\documentclass[twocolumn,superscriptaddress,amsmath,amssymb,aps,prl,10pt]{revtex4-2}

\usepackage{graphicx}
\usepackage{dcolumn}
\usepackage{bm}
\usepackage{amsthm}
\usepackage{braket}
\usepackage{ulem}
\usepackage{verbatim}
\newtheorem*{theorem*}{Theorem}

\usepackage[colorlinks,urlcolor=blue,citecolor=blue,linkcolor=blue]{hyperref}

\newcommand{\beginsupplement}{%
  \setcounter{equation}{0}\renewcommand{\theequation}{S\arabic{equation}}%
  \setcounter{figure}{0}\renewcommand{\thefigure}{S\arabic{figure}}%
  \setcounter{table}{0}\renewcommand{\thetable}{S\arabic{table}}%
  \setcounter{secnumdepth}{3}%
  \setcounter{section}{0}\renewcommand{\thesection}{\Roman{section}}%
  \setcounter{subsection}{0}\renewcommand{\thesubsection}{S\arabic{section}.\arabic{subsection}}%

  \renewcommand{\theHequation}{S\arabic{equation}}
  \renewcommand{\theHfigure}{S\arabic{figure}}
  \renewcommand{\theHtable}{S\arabic{table}}
  \renewcommand{\theHsection}{S\arabic{section}}
  \renewcommand{\theHsubsection}{S\arabic{section}.\arabic{subsection}}
}

\begin{document}
	\title{Bounds on quantum Fisher information and uncertainty relations for thermodynamically conjugate variables}

	\author{Ye-Ming Meng}
	\affiliation{State Key Laboratory of Precision Spectroscopy, Institute of Quantum Science and Precision Measurement, East China Normal University, Shanghai 200062, China}
	\author{Zhe-Yu Shi}
	\email{zyshi@lps.ecnu.edu.cn}
	\affiliation{State Key Laboratory of Precision Spectroscopy, Institute of Quantum Science and Precision Measurement, East China Normal University, Shanghai 200062, China}
	\date{\today}
	
	\begin{abstract}
    Uncertainty relations represent a foundational principle in quantum mechanics, imposing inherent limits on the precision with which \textit{mechanically} conjugate variables such as position and momentum can be simultaneously determined. This work establishes analogous relations for \textit{thermodynamically} conjugate variables --- specifically, a classical intensive parameter $\theta$ and its corresponding extensive quantum operator $\hat{O}$ --- in equilibrium states. We develop a framework to derive a rigorous thermodynamic uncertainty relation for such pairs, where the uncertainty of the classical parameter $\theta$ is quantified by its quantum Fisher information $\mathcal{F}_\theta$. The framework is based on an exact integral representation that relates $\mathcal{F}_{\theta}$ to the autocorrelation function of operator $\hat{O}$. From this representation, we derive a tight upper bound for the quantum Fisher information, which yields a thermodynamic uncertainty relation: $\Delta\theta\,\overline{\Delta O} \ge k_\text{B}T$ with $\overline{\Delta O}\equiv\partial_\theta\langle\hat{O}\rangle\,\Delta\theta$ and $T$ is the system temperature. The result establishes a fundamental precision limit for quantum sensing and metrology in thermal systems, directly connecting it to the thermodynamic properties of linear response and fluctuations.
	\end{abstract}

	\maketitle

Uncertainty relations constitute a fundamental cornerstone of quantum mechanics. They impose intrinsic limits on the precision with which multiple non-commuting observables can be simultaneously determined. The canonical formulation is the Heisenberg-Robertson uncertainty relation, which applies to any pair of Hermitian operators~\citep{Heisenberg1927,Kennard1927,Weyl1927,Robertson1929}. For \textit{mechanically} conjugate variables such as position $x$ and momentum $p$, it yields Heisenberg's well-known inequality $\Delta x\Delta p\ge\frac{\hbar}{2}$~\citep{Kennard1927,Weyl1927,Hardy1933}. The framework naturally extends to other conjugate pairs like angle and angular momentum~\footnote{
    It is worth noting that because of the compact nature of the eigenvalue
    of an angular variable, the angle-angular momentum uncertainty relation is more
    subtle than the usual position-momentum uncertainty relation, even though the
    two pairs share a similar commutation relation. Yet, it is still possible to
    derive a generalized uncertainty relation for a pair of conjugate angle and
    angular momentum by considering the uncertainty of $f(\hat{\theta})$ instead of
    $\hat{\theta}$, where $f$ is some continuous periodic function with period $2\pi$.
    See Ref.~\citep{Louisell1963,Carruthers1968} for more details.
},
the phase and the particle number of a Bose-Einstein condensate~\citep{Pethick2012}.

Notably, the concept of conjugate pairs extends beyond the field of (quantum) mechanics. For instance, \textit{thermodynamically} conjugate pairs emerge naturally in the study of thermodynamic potentials for equilibrium systems~\cite{Landau1980}. Thermodynamically conjugate quantities, such as the magnetization $M$ and magnetic field $h$ in a spin system, manifest properties analogous to those of mechanically conjugate quantities like $x$ and $p$. Specifically, both $(M,h)$ and $(x,p)$ are related by the Legendre transformations of their corresponding thermodynamic potential and Lagrangian/Hamiltonian.

This work aims to shed light on the question of whether the quantum mechanical uncertainty relation can also be extended to thermodynamically conjugate pairs. The primary difficulty in such a generalization arises from the inapplicability of the Heisenberg-Robertson formalism to thermodynamic conjugate variables. It stems from a fundamental conceptual distinction: one such variable (e.g., magnetic field $h$) typically represents a classical intensive quantity, whereas its conjugate counterpart (e.g., magnetization $M$) constitutes the expectation value of an extensive operator. Consequently, even when considering a quantum many-body system, the definition of uncertainty or fluctuation for the classical intensive quantity remains unclear.

Many attempts have been made to address this conceptual challenge. Previous research has drawn upon thermodynamic fluctuation theory~\citep{Landau1980,Cohen1979,Schlogl1988,Lindhard1986} to characterize statistical variations in thermodynamic quantities. More recently, an information-theoretic framework~\citep{Mandelbrot1962,Lavenda1987,Lavenda1988,Lavenda1988a,Lavenda1992,Uffink1999,Stace2010,Paris2015,Miller2018,Mok2021,Mehboudi2022} has been developed, which provides the conceptual basis for the present work. The framework adheres to the principle analogous to the original one proposed by Heisenberg. Specifically, despite the fixed nature of the classical intensive quantity, its experimental measurement inherently introduces uncertainty through two mechanisms: the statistical fluctuation inherent in the thermal ensemble, and the quantum mechanical uncertainty associated with the measurement itself. The measurement uncertainty of the classical quantity --- denoted as $\theta$ hereafter --- can be quantified by the variance of its estimator $\Delta\theta^2$, which is governed by the quantum Cram\'{e}r-Rao inequality~\citep{Rao1992, Helstrom1969, PARIS2009, note:CRbound-single-shot},
\begin{equation}
\Delta\theta^{2}\ge \mathcal{F}_\theta^{-1}.\label{eq:CR-bound}
\end{equation}
Here, $\mathcal{F}_\theta$ is the so-called quantum Fisher information of the system and can be uniquely determined by system's density matrix $\hat{\rho}_\theta$.

Quantum Fisher information has been extensively investigated in the context of precision measurement~\citep{Boixo2007,Pang2014,Liu2015,Froewis2015,Correa2015,Toth2022,Hauke2016}, wherein the measured (classical) quantity $\theta$ is incorporated into the out-of-equilibrium evolution of a density matrix parameterized by $\theta$. While in this work, as we focus on thermal equilibrium systems, the parameter $\theta$ is encoded in the equilibrium density matrix through a $\theta$-dependent Hamiltonian $\hat{H}(\theta)$. In the following, we consider a Gibbs ensemble with inverse temperature $\beta$, as in Ref.~\citep{GarciaPintos2024, Abiuso2025}
\begin{equation}
    \hat{\rho}_\theta=\frac{e^{-\beta \hat{H}(\theta)}}{\text{Tr}[e^{-\beta \hat{H}(\theta)}]}.\label{eq:Gibbs}
\end{equation}

The thermodynamic variable that conjugates to parameter $\theta$ is thus the thermal average (expectation value) of quantum operator $\hat{O}\equiv \partial_\theta \hat{H}(\theta)$~\citep{Landau1980}. Interestingly, it has been proved that the variance of the conjugate operator $\hat{O}$ gives a natural upper bound on the quantum Fisher information $\mathcal{F}_\theta$, i.e., $\mathcal{F}_\theta \le \beta^2 \braket{(\Delta\hat{O})^2}$, where $\Delta\hat{O}=\hat{O}-\braket{\hat{O}}$ and the brackets denote the combined quantum and thermal average, $\braket{\cdot} = \text{Tr}[\hat{\rho}_\theta(\cdot)]$~\citep{GarciaPintos2024, Abiuso2025}. The upper bound together with the Cram\'er-Rao inequality in Eq.~\eqref{eq:CR-bound} naturally leads to a thermodynamic uncertainty relation,
\begin{equation}
    \Delta\theta\Delta O \ge 1/\beta, \label{eq:TUR-variance}
\end{equation}
where we have defined the observable's standard deviation as $\Delta O\equiv\sqrt{\braket{(\Delta\hat{O})^2}}$.

In this work, we establish a new framework for deriving (tighter) upper- and lower-bounds of the quantum Fisher information $\mathcal{F}_\theta$ by relating it to the fluctuation spectrum of the conjugate operator $\hat{O}$. The foundation of this framework is an exact integral representation for the quantum Fisher information,
\begin{equation} \label{eq:QFI_fluctuation}
    \mathcal{F}_{\theta}=\frac{2}{\pi}\int_{-\infty}^{+\infty}d\omega\tanh^{2}(\frac{\beta\omega}{2})\frac{1}{\omega^{2}}S(\omega),
\end{equation}
where $S(\omega)$ is the autocorrelation function of the conjugate observable $\hat{O}$.
The formula links the metrological uncertainty in the classical thermal variable $\theta$ (quantified by $\mathcal{F}_\theta$) with the intrinsic fluctuations spectrum ($S(\omega)$) of its conjugate observable $\hat{O}$. By relaxing the integral kernel of this formula, we derive a new chain of inequalities that universally bounds $\mathcal{F}_\theta$, i.e.,
\begin{equation}
    \frac{(\partial_{\theta}\braket{\hat{O}})^{2}}{\braket{(\Delta\hat{O})^{2}}}\le\mathcal{F}_{\theta}\le\beta\partial_{\theta}\braket{\hat{O}}\le\beta^{2}\braket{(\Delta\hat{O})^{2}}. \label{eq:QFI-bounds}
\end{equation}

\begin{figure}[t]
  \centering
  \includegraphics[width=0.98\linewidth]{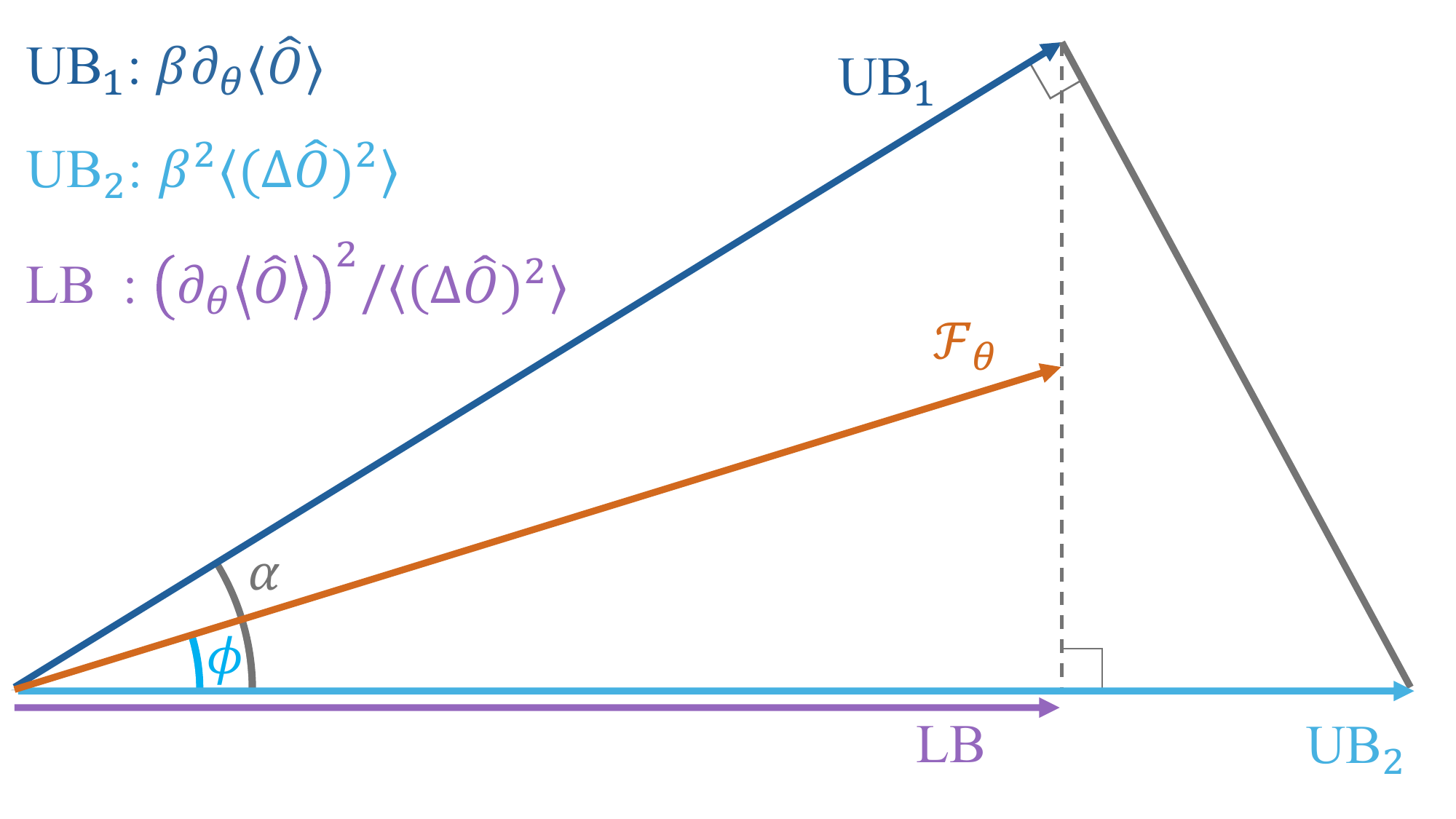}\\

  \caption{Schematic illustration of the chain inequality $\text{LB}\le\mathcal{F}_{\theta}\le\text{UB}_{1}\le\text{UB}_{2}$. The diagram highlights that these bounds are not independent, the tighter upper bound (UB$_1$) is the geometric mean of the lower bound (LB) and the conventional variance bound (UB$_2$), satisfying $\text{UB}_{1}^{2}=\text{UB}_{2}\times\text{LB}$.} \label{fig:triangular}
\end{figure}

Here, we highlight three significant features of this chain of inequalities. First, the last inequality recovers the previously mentioned variance upper bound for the quantum Fisher information proved in Refs.~\citep{GarciaPintos2024, Abiuso2025}.
Second, it introduces a new upper bound related to the thermodynamic susceptibility $\partial_\theta \braket{\hat{O}}$, which provides a strictly tighter constraint than the variance bound. Third, the two upper bounds and the lower bound of $\mathcal{F}_\theta$ (respectively denoted by $\text{UB}_2,\ \text{UB}_1,\ \text{LB}$ in the descending order hereafter) are connected by the following equality,
$\text{UB}_{1}^{2}=\text{UB}_{2}\times\text{LB}$.
It indicates that $\text{UB}_1$ is the geometric mean of $\text{UB}_2$ and $\text{LB}$, and allows us to represent Eq.~\eqref{eq:QFI-bounds} geometrically as illustrated in Fig.~\ref{fig:triangular}.

In the following, we first present the proof of the main formula represented in Eq.~\eqref{eq:QFI_fluctuation}. The approach involves establishing a relationship between the quantum Fisher information $\mathcal{F}_\theta$ of an equilibrium system at and the imaginary part of its response function $\chi_\theta$ (i.e., the dissipation) when perturbed by the classical variable $\theta$. Crucially, for thermally equilibrium systems, the dissipation of $\theta$ can be related to the fluctuation spectrum of the conjugate operator $\hat{O}$ through the fluctuation-dissipation theorem, which proves Eq.~\eqref{eq:QFI_fluctuation}. Building on this result, we establish both upper and lower bounds for $\mathcal{F}_\theta$ (Eq.~\eqref{eq:QFI-bounds}) and examine the resulting thermodynamic uncertainty relations. The theoretical framework is subsequently tested through numerical simulations using an equilibrium system of a one-dimensional spin chain undergoing a quantum phase transition.

\textit{Proof of Eq.~(\ref{eq:QFI_fluctuation}).} We start with the formula of the quantum Fisher information $\mathcal{F}_\theta$ of a general density matrix $\hat{\rho}_\theta$~\citep{Zanardi2007, PARIS2009, Liu2019}
\begin{equation}\label{eq:QFI0}
    \mathcal{F}_\theta = \sum_n \frac{(\partial_\theta p_n)^2}{p_n}+\sum_{m \neq n} \frac{2(p_m-p_n)^2}{p_m+p_n} \vert \braket{n|\partial_\theta m} \vert^2,
\end{equation}
where $|n\rangle$ and $p_n$ are eigenstate and eigenvalue of $\hat{\rho}_{\theta}$. 

For a Gibbs ensemble as specified in Eq.~\eqref{eq:Gibbs}, the density matrix can be diagonalized simultaneously with the Hamiltonian, hence $|n\rangle$ is an eigenstate of $\hat{H}$ and $p_n=e^{-\beta E_n}/\sum_ne^{-\beta E_n}$ with $E_n$ being the eigen-energy of $|n\rangle$. As previously established, when the Hamiltonian is parameterized by a classical quantity $\theta$, $(\theta,\hat{O})=(\theta,\partial_\theta\hat{H})$ constitute a pair of thermodynamically conjugate variables. Several established examples include the chemical potential and particle number $(\mu, \hat{N})$, the magnetic field and total magnetization $(h, \hat{M})$, the inverse scattering length and Tan’s contact $(1/a_s, \hat{C})$, and the squared relative velocity and superfluid density $(w^2, \hat{\rho}_s)$~\citep{Landau1980, Tan2008, Chen_2014, Liu2019}. We note that in each of these pairs, the first component represents a fixed intensive classical parameter, while the other corresponds to a fluctuating extensive quantum operator.

Under the above setup, Eq.~\eqref{eq:QFI0} can be simplified by utilizing the Hellmann–Feynman theorem and the second-order perturbation theory~\citep{note:Supplemental}, which leads to
\begin{align} \label{eq:QFI1}
    \mathcal{F}_\theta=& \beta^2 \sum_n p_n(O_{nn}-\braket{\hat{O}})^2 \nonumber \\
&+ 2 \sum_{E_m \neq E_n}\frac{(p_m-p_n)^2}{p_m+p_n} \frac{1}{(E_m-E_n)^2} |O_{mn}|^2
\end{align}
with $O_{mn}\equiv\braket{m|\hat{O}|n}$ being the matrix element of $\hat{O}$ (in the basis of $\hat{H}$). We note that the expression for the quantum Fisher information in Eq.~(\ref{eq:QFI1}) naturally separates into two distinct contributions.
The first term admits an alternative form, i.e., $\sum_n {(\partial_\theta p_n)^2}/{p_n}$, whose nature is purely statistical under the interpretation of $\{p_n\}$ as a classical distribution
\footnote{This classical contribution alone is sufficient to establish a classical thermodynamic uncertainty relations, such as Mandelbrot's bound on the temperature-energy uncertainty~\citep{Mandelbrot1962}.}. In contrast, the second term is purely quantum in origin, arising from the non-commutative nature of the operator $\hat{O}$ with the Hamiltonian $\hat{H}$.

From a physical perspective, the quantum Fisher information $\mathcal{F}_\theta$ quantifies the amount of information about the thermal variable $\theta$ carried by the equilibrium ensemble $\hat{\rho}_\theta$, which implies that it is closely related to the \textit{response} of the system followed by a perturbation in $\theta$. Indeed, it can be demonstrated that $\mathcal{F}_\theta$ can be expressed in terms of the Kubo response function $\chi(\omega)$ through
\footnote{A formal analogy exists between Eq.~(\ref{eq:QFI_dissipation}) and Eq.~(4) in Ref.~\citep{Hauke2016}. However, the physical settings are distinct. Our expression pertains to a system in thermal equilibrium, whereas theirs was developed for the context of unitary encoding.}
\begin{align}\label{eq:QFI_dissipation}
\mathcal{F}_\theta=& \beta^2 \sum_n p_n(O_{nn}-\braket{\hat{O}})^2 \nonumber \\
&+ \frac{2}{\pi} \int_{-\infty}^{+\infty} d\omega \tanh(\frac{\omega \beta}{2}) \frac{1}{\omega^2} \text{Im}[\chi(\omega)].
\end{align}
The formula can be proved by noting the resemblance between Eq.~\eqref{eq:QFI1} and the Lehmann representation of the response function $\chi(\omega)$, i.e.,~\citep{Coleman2015}
\begin{align}\label{eq:kubo_response}
\chi(\omega) =\lim_{\eta\to0^{+}}\sum_{mn}\Big[&-\frac{p_{m}}{\omega+E_{m}-E_{n}+i\eta}\nonumber \\
&+\frac{p_{n}}{\omega+E_{m}-E_{n}+i\eta}\Big]\vert O_{mn}\vert^{2}.
\end{align}
Take the imaginary (i.e., the dissipation) part of the equation, one has $\text{Im}[\chi(\omega)]=\sum_{E_m \neq E_n}(p_{m}-p_{n})\pi\delta(\omega+E_{m}-E_{n})\vert O_{mn}\vert^{2}$. Substituting this expression into Eq.~(\ref{eq:QFI_dissipation}) recovers Eq.~(\ref{eq:QFI1}), thus proves Eq.~(\ref{eq:QFI_dissipation}).

To establish a thermodynamic uncertainty relation, we seek to connect the quantum Fisher information for the variable $\theta$ with a measure of uncertainty for its conjugate quantity $\hat{O}$. This can be achieved by recognizing that the integration on the right-hand side of Eq.~\eqref{eq:QFI_dissipation} represents a weighted average of the operator $\hat{O}$'s dissipative response $\text{Im}[\chi(\omega)]$. It is thus natural to apply the fluctuation-dissipation theorem, which relates the dissipative response to the fluctuation spectrum $S(\omega)$ --- the Fourier transform of the autocorrelation function $S(t)=(\braket{\Delta\hat{O}(t)\Delta\hat{O}}+\braket{\Delta\hat{O}\Delta\hat{O}(t)})/2$ --- thereby yielding the main result given in Eq.~\eqref{eq:QFI_fluctuation}~\footnote{It should be noted that a direct substitution using the Callen-Welton~\citep{Callen1951} fluctuation-dissipation relation, i.e., $S(\omega)=\coth(\beta\omega/2) \text{Im}[\chi(\omega)]$, would incorrectly introduce an additional term in Eq.~\eqref{eq:QFI_fluctuation}. This issue arises from the divergent behavior of both the integrand in Eq.~\eqref{eq:QFI_dissipation} and the fluctuation-dissipation relation itself. The treatment of this subtlety at $\omega\rightarrow0$ is provided in the supplementary material~\citep{note:Supplemental}.}

\textit{Upper and lower bounds on $\mathcal{F}_\theta$.} The integral representation in Eq.~\eqref{eq:QFI_fluctuation} serves as an effective tool for estimating the upper and lower bounds on the quantum Fisher information, which has numerous applications in quantum metrology~\citep{Giovannetti2006, Boixo2007, Hyllus2012, Liu2015, Marzolino2017, Pezze2018, Sone2021, Beckey2022, Toth2022, Ding2023} as well as in deriving various uncertainty relations~\citep{Luo2003, Gibilisco2006, Gibilisco2007, Froewis2015, Miller2018, Toth2022,Tonchev2021}.

To demonstrate the utility of the integral representation, note that $S(\omega)$ is non-negative~\footnote{This can be verified directly from the Lehmann representation $S(\omega)=\sum_{E_m\neq E_n} (p_m+p_n) |O_{mn}-\delta_{mn}\braket{\hat{O}}|^2 \pi \delta(\omega+E_m-E_n)$}, and the weight factor $\tanh^{2}(\frac{\beta\omega}{2})\frac{1}{\omega^{2}}$ in the integration satisfies the inequality
$\tanh^{2}(\frac{\beta\omega}{2})\frac{1}{\omega^{2}} \le \tanh(\frac{\beta\omega}{2})\frac{\beta}{2\omega}\le\frac{\beta^2}{4}$. This directly leads to the two upper bounds presented in Eq.~\eqref{eq:QFI-bounds} with
\begin{align}
    &\beta\partial_{\theta}\braket{\hat{O}}=\frac{2}{\pi}\int_{-\infty}^{+\infty}d\omega\tanh(\frac{\beta\omega}{2})\frac{\beta}{2\omega} S(\omega), \label{eq:ub1_integral} \\
        &\beta^{2}\braket{(\Delta\hat{O})^{2}}=\frac{2}{\pi}\int_{-\infty}^{+\infty}d\omega \frac{\beta^2}{4} S(\omega). \label{eq:O2_integral}
\end{align}

On the other hand, the lower bound of $\mathcal{F}_\theta$ can be obtained through the following Cauchy-Schwarz inequality,
\begin{align}
    &\mathcal{F}_\theta \cdot \beta^{2}\braket{(\Delta\hat{O})^{2}} \nonumber \\
    =&\left[\frac{2}{\pi}\int_{-\infty}^{+\infty}d\omega \tanh^2(\frac{\beta\omega}{2})\frac{1}{\omega^2}S(\omega)\right]
    \cdot \left[\frac{2}{\pi}\int_{-\infty}^{+\infty} \frac{\beta^2}{4}S(\omega)d\omega\right] \nonumber \\
    \ge&\left[\frac{2}{\pi}\int_{-\infty}^{+\infty}d\omega\left(\frac{1}{\omega}\tanh(\frac{\beta\omega}{2})\sqrt{S(\omega)}\right)\left(\frac{\beta}{2} \sqrt{S(\omega)}\right)\right]^2 \nonumber \\
    =&(\beta\partial_{\theta}\braket{\hat{O}})^2.
\end{align}
Rearranging the equality immediately recovers the lower bound of Eq.~(\ref{eq:QFI-bounds}),
a result first established by Holevo via non-commutative statistics~\citep{Holevo73, kholevo1974}.

\textit{Uncertainty relation.}
The tighter upper bound on $\mathcal{F}_\theta$, i.e., $\text{UB}_1=\beta\partial_\theta \braket{\hat{O}}$, translates via the Cram\'er-Rao inequality into a new thermodynamic uncertainty relation
\begin{equation}
\Delta\theta\, \overline{\Delta O} \ge 1/\beta, \label{eq:TUR-susceptibility}
\end{equation}
where we have defined the response-based uncertainty as $\overline{\Delta O} \equiv  \partial_\theta\braket{\hat{O}}\,\Delta\theta$. This uncertainty characterizes the deviation in an indirect measurement scenario. Specifically, it represents the uncertainty in the inferred value of $\braket{\hat{O}}$ obtained by measuring its conjugate quantity $\theta$, given prior knowledge of $\langle\hat{O}\rangle$ as a function of $\theta$.

A comparison of the two types of uncertainty, $\Delta O$ and $\overline{\Delta O}$,
is of considerable interest. From the lower bound of $\mathcal{F}_\theta$ in Eq.~\eqref{eq:QFI-bounds}, we obtain $\overline{\Delta O}^2\le(\mathcal{F}_\theta\Delta\theta^2)\cdot\Delta O^2$, which implies
\begin{equation}
\overline{\Delta O} \le \Delta O,
\end{equation}
once the measurement of $\theta$ is optimal, i.e., it saturates the Cram\'er-Rao bound $\Delta\theta^2=\mathcal{F}_\theta^{-1}$.

The result indicates that inferring the expectation value of observable $\hat{O}$
via an indirect measurement of its conjugate variable $\theta$ can achieve greater precision than the direct measurement of itself. Such an enhancement, however, comes at a cost: it relies critically on prior knowledge of the functional dependence $\theta \to \braket{\hat{O}}$,
which in turn requires precise information about the system’s temperature and all Hamiltonian parameters other than $\theta$.

\textit{Numerical verification.}
The bounds we have derived apply to any quantum system in thermal equilibrium. To verify these inequalities, we apply them to a canonical model that exhibits a quantum phase transition,
the one-dimensional transverse-field Ising model. The Hamiltonian is given by
\begin{equation}
    \hat{H}=\sin(\gamma) \sum_{i=1}^{N} \sigma_{i}^{z} - \cos(\gamma) \sum_{i=1}^{N-1} \sigma_{i}^{x} \sigma_{i+1}^{x}+\theta \sum_{i=1}^N\sigma_i^x.
\end{equation}
where $\sigma_i^{x/z}$ are the Pauli operators at site $i$ on a one-dimensional chain of $N$ sites.
In our analysis, we consider the estimation of a parameter $\theta$, for which the thermodynamic conjugate observable is the total magnetization, $\hat{O} = \sum_i \sigma_i^x$.
For all subsequent calculations, we take $\theta = 0$.
The parameter $\theta$ is thus introduced only as a formal device, used solely to define the quantum Fisher information,
$\mathcal{F}_\theta$, and its corresponding conjugate observable, $\hat{O}$.
Note that $\braket{\hat{O}}$ also serves as the order parameter for the system's quantum phase transition.
At zero temperature, the model exhibits a quantum phase transition at the critical point $\gamma_c=\pi/4$, marking a symmetry-breaking transition between a ferromagnetic phase with spontaneously broken $\mathbb{Z}_2$ symmetry ($\gamma<\gamma_c$) and a symmetric paramagnetic phase ($\gamma>\gamma_c$)~\citep{SCHULTZ1964,Derzhko1997,Sachdev2011}.

\begin{figure}[t]
  \centering
  \includegraphics[width=0.98\linewidth]{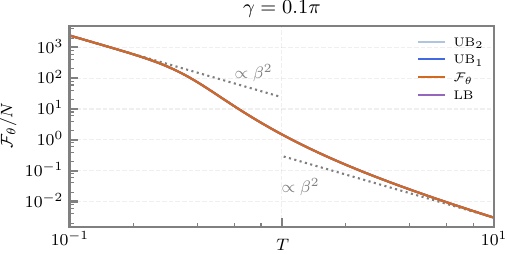}\\
  \includegraphics[width=0.98\linewidth]{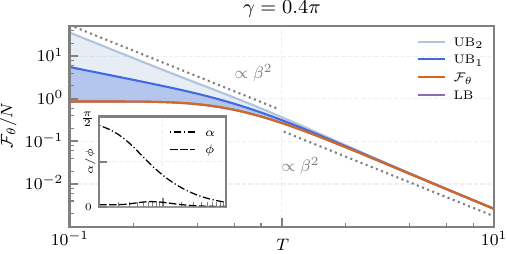}

  \caption{Temperature dependence of the quantum Fisher information (solid orange) and its bounds --- LB (purple), UB$_1$ (deep blue), and UB$_2$ (light blue). Results are for the exact solution with system size $N=100$.
    \textbf{(a)} In the ferromagnetic phase ($\gamma < \gamma_c$), all bounds are degenerate. 
    \textbf{(b)} In the paramagnetic phase ($\gamma > \gamma_c$), they are degenerate only at high $T$. In contrast, at low $T$, the quantum Fisher information is saturated by the LB. 
    The dashed line indicates the $1/T^2$ scaling followed by the quantum Fisher information at high $T$ and in the low $T$ ferromagnetic phase, highlighting cooling as a metrological resource.\\
    }\label{fig:qfi_vs_T}
\end{figure}

\begin{figure}[t]
  \centering
  \includegraphics[width=0.97\linewidth]{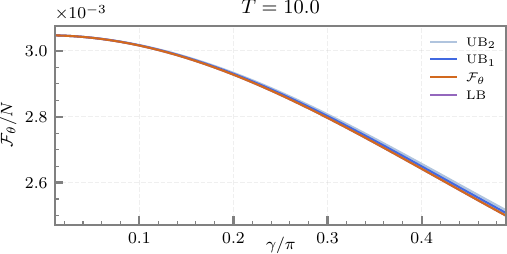}\\[1mm]
  \includegraphics[width=0.97\linewidth]{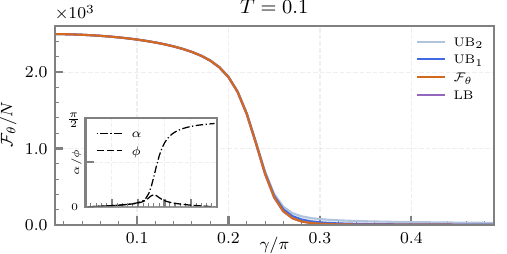}

  \caption{Transverse field dependence of the quantum Fisher information (solid orange) and its bounds --- LB (purple), UB$_1$ (deep blue), and UB$_2$ (light blue).
  Results are for the exact free-fermion solution with system size $N=100$.
  \textbf{(a)} At high temperature, all bounds are degenerate across the entire range of field strength $\gamma$.
  \textbf{(b)} At low temperature, the system exhibits dramatically different behavior on either side of the quantum phase transition point at $\gamma_c$.
  Throughout the entire ferromagnetic phase ($\gamma < \gamma_c$), the quantum Fisher information is significantly enhanced, while in the paramagnetic phase ($\gamma > \gamma_c$) the quantum Fisher information is tightly tracked by the LB.\\
  } \label{fig:qfi_vs_gamma}
\end{figure}

We first compare the relationship between the quantum Fisher information and its derived bounds across a wide range of temperatures (Fig.~\ref{fig:qfi_vs_T}).
The numerical results show that across most parameter regimes, all three bounds track the value of the quantum Fisher information closely, providing tight estimates. 
More specifically, the bounds become degenerate and collapse to a single value in two important physical limits.
The first is the high-temperature limit,
where thermal energy is much larger than the system's energy scales ($\beta\omega \ll 1 $). In this region, all bounds are identical since $ \tanh(\frac{\beta\omega}{2})\approx\frac{\beta\omega}{2}$. The second limit occurs in physical regimes where the fluctuation spectrum $S(\omega)$ is sharply peaked at zero frequency. This is the case, for instance, when the system is deep in the ferromagnetic phase with a small $\gamma$.
In this region, the low-energy sector is dominated by two nearly degenerate ferromagnetic ground states polarized along opposite directions,
which are only weakly coupled by the observable $\hat{O}$.
This leads to a slowly varying time-correlation function $S(t)$ and, consequently, a spectrum $S(\omega)$ that is sharply concentrated near zero frequency. Consequently, the integral of Eq.~\eqref{eq:QFI_fluctuation} is dominated by its low-frequency part, where $\frac{1}{\omega}\tanh(\frac{\beta\omega}{2})\approx\beta/2$ and all the bounds collapse again.

Remarkably, in both cases, all bounds collapse onto a single curve that exhibits a characteristic $1/T^2$ asymptotic scaling in both high and low-temperature limits, which is illustrated in Fig.~\ref{fig:qfi_vs_T}. The behavior can be understood by considering the asymptotic behavior of $\braket{(\Delta \hat{O})^2}$ at the temperature extremes. In the high-temperature limit, the variance can be expanded in powers of $\beta$, with the leading term $\text{Tr}[(\Delta \hat{O})^2]$ a constant value.
While in the low temperature limit, the variance approaches the ground-state variance, given by $\text{Tr}[P_{\text{GS}}(\Delta \hat{O})^2]/g_{\text{GS}}$, where $g_{\text{GS}}$ and $P_{\text{GS}}$ are the degeneracy and projector to the ground state subspace. Consequently, the asymptotic scaling of $\text{UB}_2\equiv\beta^2\langle\Delta\hat{O}^2\rangle$ is governed entirely by the $\beta^2$ prefactor.

Fig~\ref{fig:qfi_vs_gamma} illustrates the behavior of the quantum Fisher information as a function of the field-strength parameter $\gamma$.
At high temperatures (e.g., $T=10$), the different bounds are nearly degenerate across the entire range of $\gamma$, consistent with the previous analysis. The low-temperature behavior (e.g., $T=0.1$), however, is much richer and reveals the impact of the quantum phase transition, with two distinct physical regimes.

At the critical region ($\gamma \approx \gamma_c$) and throughout the paramagnetic phase ($\gamma > \gamma_c$), the bounds exhibit clear separation.
Within this regime, the upper bound $\text{UB}_1\equiv\beta\partial_{\theta}\braket{\hat{O}}$ provides a substantially tighter constraint on the quantum Fisher information compared to $\text{UB}_2$. This is evident from the quantum Fisher information shown in Fig.~\ref{fig:qfi_vs_gamma}, as well as from the inset which displays the $\alpha\rightarrow\pi/2$ behavior in the paramagnetic phase (noting that $\text{UB}_1=\text{UB}_2\cos\alpha$ as indicated in Fig.~\ref{fig:triangular}). Furthermore, the consistently small angle $\phi$ across the entire parameter space demonstrates that the lower bound $\text{LB}=\text{UB}_1^2/\text{UB}2=\mathcal{F}_\theta\cos\phi$ serves as an accurate approximation for $\mathcal{F}_\theta$ throughout this region.

In the ferromagnetic phase ($\gamma < \gamma_c$), a distinct jump in $\mathcal{F}_\theta$ is observed in the zero-temperature limit. This jump indicates enhanced sensitivity of the system to variations in the parameter $\theta$ within the ferromagnetic phase. The underlying mechanism is the spontaneous $Z_2$ symmetry breaking that defines the ferromagnetic phase. In this regime, the ground state exhibits degeneracy, and an infinitesimal longitudinal field $\theta$ suffices to break this symmetry, resulting in a substantial modification of the system's state and consequently a large quantum Fisher information that scales as $1/T^2$. In contrast, the paramagnetic phase exhibits lifted degeneracy, which leads to a considerably smaller $\mathcal{F}_\theta$, reflecting the system's robustness to the longitudinal field $\theta$. This low-temperature behavior establishes that ground-state degeneracy can serve as a quantum resource, enabling a $1/T^2$ scaling that enhances measurement precision --- a feature anticipated to be characteristic of a broad class of many-body systems.

\textit{Experimental achievability.} It is important to note that while the numerical results confirm the validity of the bounds on the quantum Fisher information $\mathcal{F}_\theta$, the tightness and practical utility of the thermodynamic uncertainty relation in Eq.~\eqref{eq:TUR-susceptibility} depend on the experimental protocol employed. In particular, if the estimation or measurement of the parameter $\theta$ yields a variance significantly larger than the lower bound prescribed by the Cram\'{e}r-Rao inequality, i.e., $\Delta\theta\gg\mathcal{F}_\theta^{-1/2}$, the uncertainty relation presented in this work would provide limited practical value. In other words, the effectiveness of our thermodynamic uncertainty relation requires that a (close to) optimal estimator of $\theta$ be experimentally accessible.

It is known that the optimal estimator of $\theta$ that reaches the Cram\'{e}r-Rao bound is given by~\citep{Braunstein1994}
\begin{align}
    \hat{\theta} = \theta + \frac{\hat{L}}{\text{Tr}[\rho_\theta \hat{L}^2]},
\end{align}
where $\hat{L}$ denotes the symmetric logarithmic derivative that satisfies $\partial_\theta \rho_\theta = \frac{1}{2}(\rho_\theta \hat{L}+\hat{L} \rho_\theta)$.

To ensure experimental accessibility of $\hat{\theta}$, it is generally necessary for it to be well-approximated by a sum of \textit{local} operators. In the supplementary material~\citep{note:Supplemental}, we prove that when both the Hamiltonian $\hat{H}$ and the operator $\hat{O}$ are sums of some local operators, and the system's correlation functions can be controlled by the Lieb-Robinson bound~\citep{Lieb1972,Nachtergaele2006}, the operator $\hat{L}$ can indeed be well-approximated by a sum of local operators. The proof relies on a new integral representation of the logarithmic derivative for the Gibbs ensemble, which is of interest in its own right,
\begin{align}\label{eq:SLD_integral}
\hat{L}=\frac{2}{\pi}\int_{-\infty}^{+\infty} dt  \log\left[\tanh(\frac{\pi|t|}{2\beta})\right] \hat{O}(t).
\end{align}
The prefactor $\log\left[\tanh(\frac{\pi|t|}{2\beta})\right]$ in the integrand exhibits exponential decay for $t\gg\beta$, which implies that $\hat{L}$ can be accurately approximated by a weighted average of $\hat{O}(t)$ over times $t\lesssim\beta$. Consequently, it is natural to expect that both the symmetric logarithmic derivative $\hat{L}$ and hence the optimal estimator $\hat{\theta}$ can be well-approximated by sums of local operators, rendering them amenable to effective experimental measurement.

\textit{Conclusions.}
We have established a new framework for systematically deriving bounds on the quantum Fisher information in thermal systems.
This framework is founded upon an exact integral representation Eq.~(\ref{eq:QFI_fluctuation}) that formally connects the quantum Fisher information, $\mathcal{F}_{\theta}$, of an intensive parameter $\theta$ to the fluctuation spectrum, $S(\omega)$, of its extensive conjugate operator, $\hat{O}$. From this integral form, a hierarchical chain of inequalities Eq.~\eqref{eq:QFI-bounds} is derived, linking the quantum Fisher information to the system's core thermodynamic properties: its linear response (susceptibility, $\partial_{\theta}\langle\hat{O}\rangle$) and its equilibrium fluctuations (variance, $\langle(\Delta\hat{O})^2\rangle$). This chain includes a novel upper bound, $\mathcal{F}_{\theta} \le \beta\partial_{\theta}\langle\hat{O}\rangle$, which is demonstrably more stringent than the previously proved bound $\langle\Delta \hat{O}^2\rangle$. This new bound, in conjunction with the Cram\'{e}r-Rao inequality, yields a new, tighter thermodynamic uncertainty relation that fundamentally constrains the precision of a parameter estimate by the system's susceptibility. Numerical validation using the 1D transverse-field Ising model confirmed the new bound's utility, particularly in the paramagnetic phase near the quantum critical point. Furthermore, the optimal estimators can be well-approximated by sums of local operators. These findings establish a new class of uncertainty relations for thermodynamic conjugate variables, revealing that the product of the uncertainty in an intensive parameter and its extensive conjugate is fundamentally bounded by the inverse temperature.

\begin{acknowledgements}
\textit{Acknowledgements.}
We are grateful for the helpful discussions with Xingze Qiu. The work is supported by the National Natural Science Foundation of China (Grant Nos. 12574293, 92576205).
\end{acknowledgements}

\bibliographystyle{apsrev4-2}
\bibliography{refs}

\begin{thebibliography}{75}%
\makeatletter
\providecommand \@ifxundefined [1]{%
 \@ifx{#1\undefined}
}%
\providecommand \@ifnum [1]{%
 \ifnum #1\expandafter \@firstoftwo
 \else \expandafter \@secondoftwo
 \fi
}%
\providecommand \@ifx [1]{%
 \ifx #1\expandafter \@firstoftwo
 \else \expandafter \@secondoftwo
 \fi
}%
\providecommand \natexlab [1]{#1}%
\providecommand \enquote  [1]{``#1''}%
\providecommand \bibnamefont  [1]{#1}%
\providecommand \bibfnamefont [1]{#1}%
\providecommand \citenamefont [1]{#1}%
\providecommand \href@noop [0]{\@secondoftwo}%
\providecommand \href [0]{\begingroup \@sanitize@url \@href}%
\providecommand \@href[1]{\@@startlink{#1}\@@href}%
\providecommand \@@href[1]{\endgroup#1\@@endlink}%
\providecommand \@sanitize@url [0]{\catcode `\\12\catcode `\$12\catcode
  `\&12\catcode `\#12\catcode `\^12\catcode `\_12\catcode `\%12\relax}%
\providecommand \@@startlink[1]{}%
\providecommand \@@endlink[0]{}%
\providecommand \url  [0]{\begingroup\@sanitize@url \@url }%
\providecommand \@url [1]{\endgroup\@href {#1}{\urlprefix }}%
\providecommand \urlprefix  [0]{URL }%
\providecommand \Eprint [0]{\href }%
\providecommand \doibase [0]{https://doi.org/}%
\providecommand \selectlanguage [0]{\@gobble}%
\providecommand \bibinfo  [0]{\@secondoftwo}%
\providecommand \bibfield  [0]{\@secondoftwo}%
\providecommand \translation [1]{[#1]}%
\providecommand \BibitemOpen [0]{}%
\providecommand \bibitemStop [0]{}%
\providecommand \bibitemNoStop [0]{.\EOS\space}%
\providecommand \EOS [0]{\spacefactor3000\relax}%
\providecommand \BibitemShut  [1]{\csname bibitem#1\endcsname}%
\let\auto@bib@innerbib\@empty
\bibitem [{\citenamefont {Heisenberg}(1927)}]{Heisenberg1927}%
  \BibitemOpen
  \bibfield  {author} {\bibinfo {author} {\bibfnamefont {W.}~\bibnamefont
  {Heisenberg}},\ }\href {https://doi.org/10.1007/bf01397280} {\bibfield
  {journal} {\bibinfo  {journal} {Zeitschrift f{\"u}r Physik}\ }\textbf
  {\bibinfo {volume} {43}},\ \bibinfo {pages} {172} (\bibinfo {year}
  {1927})}\BibitemShut {NoStop}%
\bibitem [{\citenamefont {Kennard}(1927)}]{Kennard1927}%
  \BibitemOpen
  \bibfield  {author} {\bibinfo {author} {\bibfnamefont {E.~H.}\ \bibnamefont
  {Kennard}},\ }\href {https://doi.org/10.1007/bf01391200} {\bibfield
  {journal} {\bibinfo  {journal} {Zeitschrift f{\"u}r Physik}\ }\textbf
  {\bibinfo {volume} {44}},\ \bibinfo {pages} {326} (\bibinfo {year}
  {1927})}\BibitemShut {NoStop}%
\bibitem [{\citenamefont {Weyl}(1927)}]{Weyl1927}%
  \BibitemOpen
  \bibfield  {author} {\bibinfo {author} {\bibfnamefont {H.}~\bibnamefont
  {Weyl}},\ }\href {https://doi.org/10.1007/bf02055756} {\bibfield  {journal}
  {\bibinfo  {journal} {Zeitschrift f{\"u}r Physik}\ }\textbf {\bibinfo
  {volume} {46}},\ \bibinfo {pages} {1} (\bibinfo {year} {1927})}\BibitemShut
  {NoStop}%
\bibitem [{\citenamefont {Robertson}(1929)}]{Robertson1929}%
  \BibitemOpen
  \bibfield  {author} {\bibinfo {author} {\bibfnamefont {H.~P.}\ \bibnamefont
  {Robertson}},\ }\href {https://doi.org/10.1103/physrev.34.163} {\bibfield
  {journal} {\bibinfo  {journal} {Physical Review}\ }\textbf {\bibinfo {volume}
  {34}},\ \bibinfo {pages} {163} (\bibinfo {year} {1929})}\BibitemShut
  {NoStop}%
\bibitem [{\citenamefont {Hardy}(1933)}]{Hardy1933}%
  \BibitemOpen
  \bibfield  {author} {\bibinfo {author} {\bibfnamefont {G.~H.}\ \bibnamefont
  {Hardy}},\ }\href {https://doi.org/10.1112/jlms/s1-8.3.227} {\bibfield
  {journal} {\bibinfo  {journal} {Journal of the London Mathematical Society}\
  }\textbf {\bibinfo {volume} {s1-8}},\ \bibinfo {pages} {227} (\bibinfo {year}
  {1933})}\BibitemShut {NoStop}%
\bibitem [{Note1()}]{Note1}%
  \BibitemOpen
  \bibinfo {note} {It is worth noting that because of the compact nature of the
  eigenvalue of an angular variable, the angle-angular momentum uncertainty
  relation is more subtle than the usual position-momentum uncertainty
  relation, even though the two pairs share a similar commutation relation.
  Yet, it is still possible to derive a generalized uncertainty relation for a
  pair of conjugate angle and angular momentum by considering the uncertainty
  of $f(\protect \hat {\theta })$ instead of $\protect \hat {\theta }$, where
  $f$ is some continuous periodic function with period $2\pi $. See
  Ref.~\protect \citep {Louisell1963,Carruthers1968} for more
  details.}\BibitemShut {Stop}%
\bibitem [{\citenamefont {Pethick}\ and\ \citenamefont
  {Smith}(2012)}]{Pethick2012}%
  \BibitemOpen
  \bibfield  {author} {\bibinfo {author} {\bibfnamefont {C.~J.}\ \bibnamefont
  {Pethick}}\ and\ \bibinfo {author} {\bibfnamefont {H.}~\bibnamefont
  {Smith}},\ }\href@noop {} {\emph {\bibinfo {title} {Bose-Einstein
  Condensation in Dilute Gases}}}\ (\bibinfo  {publisher} {Cambridge University
  Press},\ \bibinfo {year} {2012})\BibitemShut {NoStop}%
\bibitem [{\citenamefont {Landau}\ \emph {et~al.}(1980)\citenamefont {Landau},
  \citenamefont {Lifshitz},\ and\ \citenamefont {Pitaevskii}}]{Landau1980}%
  \BibitemOpen
  \bibfield  {author} {\bibinfo {author} {\bibfnamefont {L.~D.}\ \bibnamefont
  {Landau}}, \bibinfo {author} {\bibfnamefont {E.~M.}\ \bibnamefont
  {Lifshitz}},\ and\ \bibinfo {author} {\bibfnamefont {L.~P.}\ \bibnamefont
  {Pitaevskii}},\ }\href@noop {} {\emph {\bibinfo {title} {Statistical
  physics}}},\ \bibinfo {series} {Pergamon international library of science,
  technology, engineering, and social studies}\ No.\ \bibinfo {number} {v. 5,
  9}\ (\bibinfo  {publisher} {Pergamon Press},\ \bibinfo {address} {Oxford ;
  New York},\ \bibinfo {year} {1980})\BibitemShut {NoStop}%
\bibitem [{\citenamefont {Cohen}\ and\ \citenamefont
  {Stachel}(1979)}]{Cohen1979}%
  \BibitemOpen
  \bibfield  {author} {\bibinfo {author} {\bibfnamefont {R.~S.}\ \bibnamefont
  {Cohen}}\ and\ \bibinfo {author} {\bibfnamefont {J.~J.}\ \bibnamefont
  {Stachel}},\ }\bibinfo {title} {Questions of irreversibility and ergodicity
  [1962b]},\ in\ \href {https://doi.org/10.1007/978-94-009-9349-5_56} {\emph
  {\bibinfo {booktitle} {Selected Papers of Léon Rosenfeld}}}\ (\bibinfo
  {publisher} {Springer Netherlands},\ \bibinfo {year} {1979})\ pp.\ \bibinfo
  {pages} {808--829}\BibitemShut {NoStop}%
\bibitem [{\citenamefont {Schlögl}(1988)}]{Schlogl1988}%
  \BibitemOpen
  \bibfield  {author} {\bibinfo {author} {\bibfnamefont {F.}~\bibnamefont
  {Schlögl}},\ }\href {https://doi.org/10.1016/0022-3697(88)90200-4}
  {\bibfield  {journal} {\bibinfo  {journal} {Journal of Physics and Chemistry
  of Solids}\ }\textbf {\bibinfo {volume} {49}},\ \bibinfo {pages} {679}
  (\bibinfo {year} {1988})}\BibitemShut {NoStop}%
\bibitem [{\citenamefont {Lindhard}(1986)}]{Lindhard1986}%
  \BibitemOpen
  \bibfield  {author} {\bibinfo {author} {\bibfnamefont {J.}~\bibnamefont
  {Lindhard}},\ }\href@noop {} {\bibfield  {journal} {\bibinfo  {journal} {The
  lesson of quantum theory}\ ,\ \bibinfo {pages} {99}} (\bibinfo {year}
  {1986})}\BibitemShut {NoStop}%
\bibitem [{\citenamefont {Mandelbrot}(1962)}]{Mandelbrot1962}%
  \BibitemOpen
  \bibfield  {author} {\bibinfo {author} {\bibfnamefont {B.}~\bibnamefont
  {Mandelbrot}},\ }\href@noop {} {\bibfield  {journal} {\bibinfo  {journal}
  {The Annals of Mathematical Statistics}\ ,\ \bibinfo {pages} {1021}}
  (\bibinfo {year} {1962})}\BibitemShut {NoStop}%
\bibitem [{\citenamefont {Lavenda}(1987)}]{Lavenda1987}%
  \BibitemOpen
  \bibfield  {author} {\bibinfo {author} {\bibfnamefont {B.~H.}\ \bibnamefont
  {Lavenda}},\ }\href {https://doi.org/10.1007/bf00669362} {\bibfield
  {journal} {\bibinfo  {journal} {International Journal of Theoretical
  Physics}\ }\textbf {\bibinfo {volume} {26}},\ \bibinfo {pages} {1069}
  (\bibinfo {year} {1987})}\BibitemShut {NoStop}%
\bibitem [{\citenamefont {Lavenda}(1988{\natexlab{a}})}]{Lavenda1988}%
  \BibitemOpen
  \bibfield  {author} {\bibinfo {author} {\bibfnamefont {B.~H.}\ \bibnamefont
  {Lavenda}},\ }\href {https://doi.org/10.1007/bf00669394} {\bibfield
  {journal} {\bibinfo  {journal} {International Journal of Theoretical
  Physics}\ }\textbf {\bibinfo {volume} {27}},\ \bibinfo {pages} {451}
  (\bibinfo {year} {1988}{\natexlab{a}})}\BibitemShut {NoStop}%
\bibitem [{\citenamefont {Lavenda}(1988{\natexlab{b}})}]{Lavenda1988a}%
  \BibitemOpen
  \bibfield  {author} {\bibinfo {author} {\bibfnamefont {B.}~\bibnamefont
  {Lavenda}},\ }\href {https://doi.org/10.1016/0022-3697(88)90201-6} {\bibfield
   {journal} {\bibinfo  {journal} {Journal of Physics and Chemistry of Solids}\
  }\textbf {\bibinfo {volume} {49}},\ \bibinfo {pages} {685} (\bibinfo {year}
  {1988}{\natexlab{b}})}\BibitemShut {NoStop}%
\bibitem [{\citenamefont {Lavenda}(1992)}]{Lavenda1992}%
  \BibitemOpen
  \bibfield  {author} {\bibinfo {author} {\bibfnamefont {B.~H.}\ \bibnamefont
  {Lavenda}},\ }\href@noop {} {\emph {\bibinfo {title} {Statistical physics: a
  probabilistic approach}}},\ A {Wiley}-{Interscience} publication\ (\bibinfo
  {publisher} {Wiley},\ \bibinfo {address} {New York},\ \bibinfo {year}
  {1992})\BibitemShut {NoStop}%
\bibitem [{\citenamefont {Uffink}\ and\ \citenamefont {van
  Lith}(1999)}]{Uffink1999}%
  \BibitemOpen
  \bibfield  {author} {\bibinfo {author} {\bibfnamefont {J.}~\bibnamefont
  {Uffink}}\ and\ \bibinfo {author} {\bibfnamefont {J.}~\bibnamefont {van
  Lith}},\ }\href {https://doi.org/10.1023/a:1018811305766} {\bibfield
  {journal} {\bibinfo  {journal} {Foundations of Physics}\ }\textbf {\bibinfo
  {volume} {29}},\ \bibinfo {pages} {655} (\bibinfo {year} {1999})}\BibitemShut
  {NoStop}%
\bibitem [{\citenamefont {Stace}(2010)}]{Stace2010}%
  \BibitemOpen
  \bibfield  {author} {\bibinfo {author} {\bibfnamefont {T.~M.}\ \bibnamefont
  {Stace}},\ }\href {https://doi.org/10.1103/physreva.82.011611} {\bibfield
  {journal} {\bibinfo  {journal} {Physical Review A}\ }\textbf {\bibinfo
  {volume} {82}},\ \bibinfo {pages} {011611} (\bibinfo {year}
  {2010})}\BibitemShut {NoStop}%
\bibitem [{\citenamefont {Paris}(2015)}]{Paris2015}%
  \BibitemOpen
  \bibfield  {author} {\bibinfo {author} {\bibfnamefont {M.~G.~A.}\
  \bibnamefont {Paris}},\ }\href
  {https://doi.org/10.1088/1751-8113/49/3/03lt02} {\bibfield  {journal}
  {\bibinfo  {journal} {Journal of Physics A: Mathematical and Theoretical}\
  }\textbf {\bibinfo {volume} {49}},\ \bibinfo {pages} {03LT02} (\bibinfo
  {year} {2015})}\BibitemShut {NoStop}%
\bibitem [{\citenamefont {Miller}\ and\ \citenamefont
  {Anders}(2018)}]{Miller2018}%
  \BibitemOpen
  \bibfield  {author} {\bibinfo {author} {\bibfnamefont {H.~J.~D.}\
  \bibnamefont {Miller}}\ and\ \bibinfo {author} {\bibfnamefont
  {J.}~\bibnamefont {Anders}},\ }\bibfield  {journal} {\bibinfo  {journal}
  {Nature Communications}\ }\textbf {\bibinfo {volume} {9}},\ \href
  {https://doi.org/10.1038/s41467-018-04536-7} {10.1038/s41467-018-04536-7}
  (\bibinfo {year} {2018})\BibitemShut {NoStop}%
\bibitem [{\citenamefont {Mok}\ \emph {et~al.}(2021)\citenamefont {Mok},
  \citenamefont {Bharti}, \citenamefont {Kwek},\ and\ \citenamefont
  {Bayat}}]{Mok2021}%
  \BibitemOpen
  \bibfield  {author} {\bibinfo {author} {\bibfnamefont {W.-K.}\ \bibnamefont
  {Mok}}, \bibinfo {author} {\bibfnamefont {K.}~\bibnamefont {Bharti}},
  \bibinfo {author} {\bibfnamefont {L.-C.}\ \bibnamefont {Kwek}},\ and\
  \bibinfo {author} {\bibfnamefont {A.}~\bibnamefont {Bayat}},\ }\bibfield
  {journal} {\bibinfo  {journal} {Communications Physics}\ }\textbf {\bibinfo
  {volume} {4}},\ \href {https://doi.org/10.1038/s42005-021-00572-w}
  {10.1038/s42005-021-00572-w} (\bibinfo {year} {2021})\BibitemShut {NoStop}%
\bibitem [{\citenamefont {Mehboudi}\ \emph {et~al.}(2022)\citenamefont
  {Mehboudi}, \citenamefont {Jørgensen}, \citenamefont {Seah}, \citenamefont
  {Brask}, \citenamefont {Kołodyński},\ and\ \citenamefont
  {Perarnau-Llobet}}]{Mehboudi2022}%
  \BibitemOpen
  \bibfield  {author} {\bibinfo {author} {\bibfnamefont {M.}~\bibnamefont
  {Mehboudi}}, \bibinfo {author} {\bibfnamefont {M.~R.}\ \bibnamefont
  {Jørgensen}}, \bibinfo {author} {\bibfnamefont {S.}~\bibnamefont {Seah}},
  \bibinfo {author} {\bibfnamefont {J.~B.}\ \bibnamefont {Brask}}, \bibinfo
  {author} {\bibfnamefont {J.}~\bibnamefont {Kołodyński}},\ and\ \bibinfo
  {author} {\bibfnamefont {M.}~\bibnamefont {Perarnau-Llobet}},\ }\href
  {https://doi.org/10.1103/physrevlett.128.130502} {\bibfield  {journal}
  {\bibinfo  {journal} {Physical Review Letters}\ }\textbf {\bibinfo {volume}
  {128}},\ \bibinfo {pages} {130502} (\bibinfo {year} {2022})}\BibitemShut
  {NoStop}%
\bibitem [{\citenamefont {Rao}(1992)}]{Rao1992}%
  \BibitemOpen
  \bibfield  {author} {\bibinfo {author} {\bibfnamefont {C.~R.}\ \bibnamefont
  {Rao}},\ }\bibinfo {title} {Information and the accuracy attainable in the
  estimation of statistical parameters},\ in\ \href
  {https://doi.org/10.1007/978-1-4612-0919-5_16} {\emph {\bibinfo {booktitle}
  {Breakthroughs in Statistics}}}\ (\bibinfo  {publisher} {Springer New York},\
  \bibinfo {year} {1992})\ pp.\ \bibinfo {pages} {235--247}\BibitemShut
  {NoStop}%
\bibitem [{\citenamefont {Helstrom}(1969)}]{Helstrom1969}%
  \BibitemOpen
  \bibfield  {author} {\bibinfo {author} {\bibfnamefont {C.~W.}\ \bibnamefont
  {Helstrom}},\ }\href {https://doi.org/10.1007/bf01007479} {\bibfield
  {journal} {\bibinfo  {journal} {Journal of Statistical Physics}\ }\textbf
  {\bibinfo {volume} {1}},\ \bibinfo {pages} {231} (\bibinfo {year}
  {1969})}\BibitemShut {NoStop}%
\bibitem [{\citenamefont {PARIS}(2009)}]{PARIS2009}%
  \BibitemOpen
  \bibfield  {author} {\bibinfo {author} {\bibfnamefont {M.~G.~A.}\
  \bibnamefont {PARIS}},\ }\href {https://doi.org/10.1142/s0219749909004839}
  {\bibfield  {journal} {\bibinfo  {journal} {International Journal of Quantum
  Information}\ }\textbf {\bibinfo {volume} {07}},\ \bibinfo {pages} {125}
  (\bibinfo {year} {2009})}\BibitemShut {NoStop}%
\bibitem [{not({\natexlab{a}})}]{note:CRbound-single-shot}%
  \BibitemOpen
  \bibinfo {note} {Note that the standard Cram\'er-Rao bound is expressed as
  $\Delta\theta^{2}\ge \frac{1}{n}\mathcal{F}_\theta^{-1}$, where $n$
  represents the number of measurements. In this work, we focus on the
  single-shot limit, which corresponds to setting $n=1$.}\BibitemShut {Stop}%
\bibitem [{\citenamefont {Boixo}\ \emph {et~al.}(2007)\citenamefont {Boixo},
  \citenamefont {Flammia}, \citenamefont {Caves},\ and\ \citenamefont
  {Geremia}}]{Boixo2007}%
  \BibitemOpen
  \bibfield  {author} {\bibinfo {author} {\bibfnamefont {S.}~\bibnamefont
  {Boixo}}, \bibinfo {author} {\bibfnamefont {S.~T.}\ \bibnamefont {Flammia}},
  \bibinfo {author} {\bibfnamefont {C.~M.}\ \bibnamefont {Caves}},\ and\
  \bibinfo {author} {\bibfnamefont {J.}~\bibnamefont {Geremia}},\ }\href
  {https://doi.org/10.1103/physrevlett.98.090401} {\bibfield  {journal}
  {\bibinfo  {journal} {Physical Review Letters}\ }\textbf {\bibinfo {volume}
  {98}},\ \bibinfo {pages} {090401} (\bibinfo {year} {2007})}\BibitemShut
  {NoStop}%
\bibitem [{\citenamefont {Pang}\ and\ \citenamefont {Brun}(2014)}]{Pang2014}%
  \BibitemOpen
  \bibfield  {author} {\bibinfo {author} {\bibfnamefont {S.}~\bibnamefont
  {Pang}}\ and\ \bibinfo {author} {\bibfnamefont {T.~A.}\ \bibnamefont
  {Brun}},\ }\href {https://doi.org/10.1103/physreva.90.022117} {\bibfield
  {journal} {\bibinfo  {journal} {Physical Review A}\ }\textbf {\bibinfo
  {volume} {90}},\ \bibinfo {pages} {022117} (\bibinfo {year}
  {2014})}\BibitemShut {NoStop}%
\bibitem [{\citenamefont {Liu}\ \emph {et~al.}(2015)\citenamefont {Liu},
  \citenamefont {Jing},\ and\ \citenamefont {Wang}}]{Liu2015}%
  \BibitemOpen
  \bibfield  {author} {\bibinfo {author} {\bibfnamefont {J.}~\bibnamefont
  {Liu}}, \bibinfo {author} {\bibfnamefont {X.-X.}\ \bibnamefont {Jing}},\ and\
  \bibinfo {author} {\bibfnamefont {X.}~\bibnamefont {Wang}},\ }\bibfield
  {journal} {\bibinfo  {journal} {Scientific Reports}\ }\textbf {\bibinfo
  {volume} {5}},\ \href {https://doi.org/10.1038/srep08565} {10.1038/srep08565}
  (\bibinfo {year} {2015})\BibitemShut {NoStop}%
\bibitem [{\citenamefont {Fr{\"o}wis}\ \emph {et~al.}(2015)\citenamefont
  {Fr{\"o}wis}, \citenamefont {Schmied},\ and\ \citenamefont
  {Gisin}}]{Froewis2015}%
  \BibitemOpen
  \bibfield  {author} {\bibinfo {author} {\bibfnamefont {F.}~\bibnamefont
  {Fr{\"o}wis}}, \bibinfo {author} {\bibfnamefont {R.}~\bibnamefont
  {Schmied}},\ and\ \bibinfo {author} {\bibfnamefont {N.}~\bibnamefont
  {Gisin}},\ }\href {https://doi.org/10.1103/physreva.92.012102} {\bibfield
  {journal} {\bibinfo  {journal} {Physical Review A}\ }\textbf {\bibinfo
  {volume} {92}},\ \bibinfo {pages} {012102} (\bibinfo {year}
  {2015})}\BibitemShut {NoStop}%
\bibitem [{\citenamefont {Correa}\ \emph {et~al.}(2015)\citenamefont {Correa},
  \citenamefont {Mehboudi}, \citenamefont {Adesso},\ and\ \citenamefont
  {Sanpera}}]{Correa2015}%
  \BibitemOpen
  \bibfield  {author} {\bibinfo {author} {\bibfnamefont {L.~A.}\ \bibnamefont
  {Correa}}, \bibinfo {author} {\bibfnamefont {M.}~\bibnamefont {Mehboudi}},
  \bibinfo {author} {\bibfnamefont {G.}~\bibnamefont {Adesso}},\ and\ \bibinfo
  {author} {\bibfnamefont {A.}~\bibnamefont {Sanpera}},\ }\href
  {https://doi.org/10.1103/physrevlett.114.220405} {\bibfield  {journal}
  {\bibinfo  {journal} {Physical Review Letters}\ }\textbf {\bibinfo {volume}
  {114}},\ \bibinfo {pages} {220405} (\bibinfo {year} {2015})}\BibitemShut
  {NoStop}%
\bibitem [{\citenamefont {T{\'o}th}\ and\ \citenamefont
  {Fr{\"o}wis}(2022)}]{Toth2022}%
  \BibitemOpen
  \bibfield  {author} {\bibinfo {author} {\bibfnamefont {G.}~\bibnamefont
  {T{\'o}th}}\ and\ \bibinfo {author} {\bibfnamefont {F.}~\bibnamefont
  {Fr{\"o}wis}},\ }\href {https://doi.org/10.1103/physrevresearch.4.013075}
  {\bibfield  {journal} {\bibinfo  {journal} {Physical Review Research}\
  }\textbf {\bibinfo {volume} {4}},\ \bibinfo {pages} {013075} (\bibinfo {year}
  {2022})}\BibitemShut {NoStop}%
\bibitem [{\citenamefont {Hauke}\ \emph {et~al.}(2016)\citenamefont {Hauke},
  \citenamefont {Heyl}, \citenamefont {Tagliacozzo},\ and\ \citenamefont
  {Zoller}}]{Hauke2016}%
  \BibitemOpen
  \bibfield  {author} {\bibinfo {author} {\bibfnamefont {P.}~\bibnamefont
  {Hauke}}, \bibinfo {author} {\bibfnamefont {M.}~\bibnamefont {Heyl}},
  \bibinfo {author} {\bibfnamefont {L.}~\bibnamefont {Tagliacozzo}},\ and\
  \bibinfo {author} {\bibfnamefont {P.}~\bibnamefont {Zoller}},\ }\href
  {https://doi.org/10.1038/nphys3700} {\bibfield  {journal} {\bibinfo
  {journal} {Nature Physics}\ }\textbf {\bibinfo {volume} {12}},\ \bibinfo
  {pages} {778} (\bibinfo {year} {2016})}\BibitemShut {NoStop}%
\bibitem [{\citenamefont {Garc{\'\i}a-Pintos}\ \emph
  {et~al.}(2024)\citenamefont {Garc{\'\i}a-Pintos}, \citenamefont {Bharti},
  \citenamefont {Bringewatt}, \citenamefont {Dehghani}, \citenamefont
  {Ehrenberg}, \citenamefont {Yunger~Halpern},\ and\ \citenamefont
  {Gorshkov}}]{GarciaPintos2024}%
  \BibitemOpen
  \bibfield  {author} {\bibinfo {author} {\bibfnamefont {L.~P.}\ \bibnamefont
  {Garc{\'\i}a-Pintos}}, \bibinfo {author} {\bibfnamefont {K.}~\bibnamefont
  {Bharti}}, \bibinfo {author} {\bibfnamefont {J.}~\bibnamefont {Bringewatt}},
  \bibinfo {author} {\bibfnamefont {H.}~\bibnamefont {Dehghani}}, \bibinfo
  {author} {\bibfnamefont {A.}~\bibnamefont {Ehrenberg}}, \bibinfo {author}
  {\bibfnamefont {N.}~\bibnamefont {Yunger~Halpern}},\ and\ \bibinfo {author}
  {\bibfnamefont {A.~V.}\ \bibnamefont {Gorshkov}},\ }\href
  {https://doi.org/10.1103/physrevlett.133.040802} {\bibfield  {journal}
  {\bibinfo  {journal} {Physical Review Letters}\ }\textbf {\bibinfo {volume}
  {133}},\ \bibinfo {pages} {040802} (\bibinfo {year} {2024})}\BibitemShut
  {NoStop}%
\bibitem [{\citenamefont {Abiuso}\ \emph {et~al.}(2025)\citenamefont {Abiuso},
  \citenamefont {Sekatski}, \citenamefont {Calsamiglia},\ and\ \citenamefont
  {Perarnau-Llobet}}]{Abiuso2025}%
  \BibitemOpen
  \bibfield  {author} {\bibinfo {author} {\bibfnamefont {P.}~\bibnamefont
  {Abiuso}}, \bibinfo {author} {\bibfnamefont {P.}~\bibnamefont {Sekatski}},
  \bibinfo {author} {\bibfnamefont {J.}~\bibnamefont {Calsamiglia}},\ and\
  \bibinfo {author} {\bibfnamefont {M.}~\bibnamefont {Perarnau-Llobet}},\
  }\href {https://doi.org/10.1103/physrevlett.134.010801} {\bibfield  {journal}
  {\bibinfo  {journal} {Physical Review Letters}\ }\textbf {\bibinfo {volume}
  {134}},\ \bibinfo {pages} {010801} (\bibinfo {year} {2025})}\BibitemShut
  {NoStop}%
\bibitem [{\citenamefont {Zanardi}\ \emph {et~al.}(2007)\citenamefont
  {Zanardi}, \citenamefont {Campos~Venuti},\ and\ \citenamefont
  {Giorda}}]{Zanardi2007}%
  \BibitemOpen
  \bibfield  {author} {\bibinfo {author} {\bibfnamefont {P.}~\bibnamefont
  {Zanardi}}, \bibinfo {author} {\bibfnamefont {L.}~\bibnamefont
  {Campos~Venuti}},\ and\ \bibinfo {author} {\bibfnamefont {P.}~\bibnamefont
  {Giorda}},\ }\href {https://doi.org/10.1103/physreva.76.062318} {\bibfield
  {journal} {\bibinfo  {journal} {Physical Review A}\ }\textbf {\bibinfo
  {volume} {76}},\ \bibinfo {pages} {062318} (\bibinfo {year}
  {2007})}\BibitemShut {NoStop}%
\bibitem [{\citenamefont {Liu}\ \emph {et~al.}(2019)\citenamefont {Liu},
  \citenamefont {Yuan}, \citenamefont {Lu},\ and\ \citenamefont
  {Wang}}]{Liu2019}%
  \BibitemOpen
  \bibfield  {author} {\bibinfo {author} {\bibfnamefont {J.}~\bibnamefont
  {Liu}}, \bibinfo {author} {\bibfnamefont {H.}~\bibnamefont {Yuan}}, \bibinfo
  {author} {\bibfnamefont {X.-M.}\ \bibnamefont {Lu}},\ and\ \bibinfo {author}
  {\bibfnamefont {X.}~\bibnamefont {Wang}},\ }\href
  {https://doi.org/10.1088/1751-8121/ab5d4d} {\bibfield  {journal} {\bibinfo
  {journal} {Journal of Physics A: Mathematical and Theoretical}\ }\textbf
  {\bibinfo {volume} {53}},\ \bibinfo {pages} {023001} (\bibinfo {year}
  {2019})}\BibitemShut {NoStop}%
\bibitem [{\citenamefont {Tan}(2008)}]{Tan2008}%
  \BibitemOpen
  \bibfield  {author} {\bibinfo {author} {\bibfnamefont {S.}~\bibnamefont
  {Tan}},\ }\href {https://doi.org/10.1016/j.aop.2008.03.004} {\bibfield
  {journal} {\bibinfo  {journal} {Annals of Physics}\ }\textbf {\bibinfo
  {volume} {323}},\ \bibinfo {pages} {2952} (\bibinfo {year}
  {2008})}\BibitemShut {NoStop}%
\bibitem [{\citenamefont {Chen}\ \emph {et~al.}(2014)\citenamefont {Chen},
  \citenamefont {Jiang}, \citenamefont {Guan},\ and\ \citenamefont
  {Zhou}}]{Chen_2014}%
  \BibitemOpen
  \bibfield  {author} {\bibinfo {author} {\bibfnamefont {Y.-Y.}\ \bibnamefont
  {Chen}}, \bibinfo {author} {\bibfnamefont {Y.-Z.}\ \bibnamefont {Jiang}},
  \bibinfo {author} {\bibfnamefont {X.-W.}\ \bibnamefont {Guan}},\ and\
  \bibinfo {author} {\bibfnamefont {Q.}~\bibnamefont {Zhou}},\ }\bibfield
  {journal} {\bibinfo  {journal} {Nature Communications}\ }\textbf {\bibinfo
  {volume} {5}},\ \href {https://doi.org/10.1038/ncomms6140}
  {10.1038/ncomms6140} (\bibinfo {year} {2014})\BibitemShut {NoStop}%
\bibitem [{not({\natexlab{b}})}]{note:Supplemental}%
  \BibitemOpen
  \bibinfo {note} {See Supplemental Material for additional analytical details
  and derivations.}\BibitemShut {Stop}%
\bibitem [{Note2()}]{Note2}%
  \BibitemOpen
  \bibinfo {note} {This classical contribution alone is sufficient to establish
  a classical thermodynamic uncertainty relations, such as Mandelbrot's bound
  on the temperature-energy uncertainty~\protect \citep
  {Mandelbrot1962}.}\BibitemShut {Stop}%
\bibitem [{Note3()}]{Note3}%
  \BibitemOpen
  \bibinfo {note} {A formal analogy exists between Eq.~(\ref
  {eq:QFI_dissipation}) and Eq.~(4) in Ref.~\protect \citep {Hauke2016}.
  However, the physical settings are distinct. Our expression pertains to a
  system in thermal equilibrium, whereas theirs was developed for the context
  of unitary encoding.}\BibitemShut {Stop}%
\bibitem [{\citenamefont {Coleman}(2015)}]{Coleman2015}%
  \BibitemOpen
  \bibfield  {author} {\bibinfo {author} {\bibfnamefont {P.}~\bibnamefont
  {Coleman}},\ }\href@noop {} {\emph {\bibinfo {title} {Introduction to
  many-body physics}}}\ (\bibinfo  {publisher} {Cambridge University Press},\
  \bibinfo {address} {Cambridge},\ \bibinfo {year} {2015})\BibitemShut
  {NoStop}%
\bibitem [{Note4()}]{Note4}%
  \BibitemOpen
  \bibinfo {note} {It should be noted that a direct substitution using the
  Callen-Welton~\protect \citep {Callen1951} fluctuation-dissipation relation,
  i.e., $S(\omega )=\protect \qopname \relax o{coth}(\beta \omega /2) \protect
  \text {Im}[\chi (\omega )]$, would incorrectly introduce an additional term
  in Eq.~\protect \textup {\hbox {\mathsurround \z@ \protect \normalfont
  (\ignorespaces \ref {eq:QFI_fluctuation}\unskip \@@italiccorr )}}. This issue
  arises from the divergent behavior of both the integrand in Eq.~\protect
  \textup {\hbox {\mathsurround \z@ \protect \normalfont (\ignorespaces \ref
  {eq:QFI_dissipation}\unskip \@@italiccorr )}} and the fluctuation-dissipation
  relation itself. The treatment of this subtlety at $\omega \rightarrow 0$ is
  provided in the supplementary material~\protect \citep
  {note:Supplemental}.}\BibitemShut {Stop}%
\bibitem [{\citenamefont {Giovannetti}\ \emph {et~al.}(2006)\citenamefont
  {Giovannetti}, \citenamefont {Lloyd},\ and\ \citenamefont
  {Maccone}}]{Giovannetti2006}%
  \BibitemOpen
  \bibfield  {author} {\bibinfo {author} {\bibfnamefont {V.}~\bibnamefont
  {Giovannetti}}, \bibinfo {author} {\bibfnamefont {S.}~\bibnamefont {Lloyd}},\
  and\ \bibinfo {author} {\bibfnamefont {L.}~\bibnamefont {Maccone}},\ }\href
  {https://doi.org/10.1103/physrevlett.96.010401} {\bibfield  {journal}
  {\bibinfo  {journal} {Physical Review Letters}\ }\textbf {\bibinfo {volume}
  {96}},\ \bibinfo {pages} {010401} (\bibinfo {year} {2006})}\BibitemShut
  {NoStop}%
\bibitem [{\citenamefont {Hyllus}\ \emph {et~al.}(2012)\citenamefont {Hyllus},
  \citenamefont {Laskowski}, \citenamefont {Krischek}, \citenamefont
  {Schwemmer}, \citenamefont {Wieczorek}, \citenamefont {Weinfurter},
  \citenamefont {Pezz{\'e}},\ and\ \citenamefont {Smerzi}}]{Hyllus2012}%
  \BibitemOpen
  \bibfield  {author} {\bibinfo {author} {\bibfnamefont {P.}~\bibnamefont
  {Hyllus}}, \bibinfo {author} {\bibfnamefont {W.}~\bibnamefont {Laskowski}},
  \bibinfo {author} {\bibfnamefont {R.}~\bibnamefont {Krischek}}, \bibinfo
  {author} {\bibfnamefont {C.}~\bibnamefont {Schwemmer}}, \bibinfo {author}
  {\bibfnamefont {W.}~\bibnamefont {Wieczorek}}, \bibinfo {author}
  {\bibfnamefont {H.}~\bibnamefont {Weinfurter}}, \bibinfo {author}
  {\bibfnamefont {L.}~\bibnamefont {Pezz{\'e}}},\ and\ \bibinfo {author}
  {\bibfnamefont {A.}~\bibnamefont {Smerzi}},\ }\href
  {https://doi.org/10.1103/physreva.85.022321} {\bibfield  {journal} {\bibinfo
  {journal} {Physical Review A}\ }\textbf {\bibinfo {volume} {85}},\ \bibinfo
  {pages} {022321} (\bibinfo {year} {2012})}\BibitemShut {NoStop}%
\bibitem [{\citenamefont {Marzolino}\ and\ \citenamefont
  {Prosen}(2017)}]{Marzolino2017}%
  \BibitemOpen
  \bibfield  {author} {\bibinfo {author} {\bibfnamefont {U.}~\bibnamefont
  {Marzolino}}\ and\ \bibinfo {author} {\bibfnamefont {T.}~\bibnamefont
  {Prosen}},\ }\href {https://doi.org/10.1103/physrevb.96.104402} {\bibfield
  {journal} {\bibinfo  {journal} {Physical Review B}\ }\textbf {\bibinfo
  {volume} {96}},\ \bibinfo {pages} {104402} (\bibinfo {year}
  {2017})}\BibitemShut {NoStop}%
\bibitem [{\citenamefont {Pezz\`e}\ \emph {et~al.}(2018)\citenamefont
  {Pezz\`e}, \citenamefont {Smerzi}, \citenamefont {Oberthaler}, \citenamefont
  {Schmied},\ and\ \citenamefont {Treutlein}}]{Pezze2018}%
  \BibitemOpen
  \bibfield  {author} {\bibinfo {author} {\bibfnamefont {L.}~\bibnamefont
  {Pezz\`e}}, \bibinfo {author} {\bibfnamefont {A.}~\bibnamefont {Smerzi}},
  \bibinfo {author} {\bibfnamefont {M.~K.}\ \bibnamefont {Oberthaler}},
  \bibinfo {author} {\bibfnamefont {R.}~\bibnamefont {Schmied}},\ and\ \bibinfo
  {author} {\bibfnamefont {P.}~\bibnamefont {Treutlein}},\ }\href
  {https://doi.org/10.1103/revmodphys.90.035005} {\bibfield  {journal}
  {\bibinfo  {journal} {Reviews of Modern Physics}\ }\textbf {\bibinfo {volume}
  {90}},\ \bibinfo {pages} {035005} (\bibinfo {year} {2018})}\BibitemShut
  {NoStop}%
\bibitem [{\citenamefont {Sone}\ \emph {et~al.}(2021)\citenamefont {Sone},
  \citenamefont {Cerezo}, \citenamefont {Beckey},\ and\ \citenamefont
  {Coles}}]{Sone2021}%
  \BibitemOpen
  \bibfield  {author} {\bibinfo {author} {\bibfnamefont {A.}~\bibnamefont
  {Sone}}, \bibinfo {author} {\bibfnamefont {M.}~\bibnamefont {Cerezo}},
  \bibinfo {author} {\bibfnamefont {J.~L.}\ \bibnamefont {Beckey}},\ and\
  \bibinfo {author} {\bibfnamefont {P.~J.}\ \bibnamefont {Coles}},\ }\href
  {https://doi.org/10.1103/physreva.104.062602} {\bibfield  {journal} {\bibinfo
   {journal} {Physical Review A}\ }\textbf {\bibinfo {volume} {104}},\ \bibinfo
  {pages} {062602} (\bibinfo {year} {2021})}\BibitemShut {NoStop}%
\bibitem [{\citenamefont {Beckey}\ \emph {et~al.}(2022)\citenamefont {Beckey},
  \citenamefont {Cerezo}, \citenamefont {Sone},\ and\ \citenamefont
  {Coles}}]{Beckey2022}%
  \BibitemOpen
  \bibfield  {author} {\bibinfo {author} {\bibfnamefont {J.~L.}\ \bibnamefont
  {Beckey}}, \bibinfo {author} {\bibfnamefont {M.}~\bibnamefont {Cerezo}},
  \bibinfo {author} {\bibfnamefont {A.}~\bibnamefont {Sone}},\ and\ \bibinfo
  {author} {\bibfnamefont {P.~J.}\ \bibnamefont {Coles}},\ }\href
  {https://doi.org/10.1103/physrevresearch.4.013083} {\bibfield  {journal}
  {\bibinfo  {journal} {Physical Review Research}\ }\textbf {\bibinfo {volume}
  {4}},\ \bibinfo {pages} {013083} (\bibinfo {year} {2022})}\BibitemShut
  {NoStop}%
\bibitem [{\citenamefont {Ding}\ \emph {et~al.}(2023)\citenamefont {Ding},
  \citenamefont {Wang},\ and\ \citenamefont {Chen}}]{Ding2023}%
  \BibitemOpen
  \bibfield  {author} {\bibinfo {author} {\bibfnamefont {W.}~\bibnamefont
  {Ding}}, \bibinfo {author} {\bibfnamefont {X.}~\bibnamefont {Wang}},\ and\
  \bibinfo {author} {\bibfnamefont {S.}~\bibnamefont {Chen}},\ }\href
  {https://doi.org/10.1103/physrevlett.131.160801} {\bibfield  {journal}
  {\bibinfo  {journal} {Physical Review Letters}\ }\textbf {\bibinfo {volume}
  {131}},\ \bibinfo {pages} {160801} (\bibinfo {year} {2023})}\BibitemShut
  {NoStop}%
\bibitem [{\citenamefont {Luo}(2003)}]{Luo2003}%
  \BibitemOpen
  \bibfield  {author} {\bibinfo {author} {\bibfnamefont {S.}~\bibnamefont
  {Luo}},\ }\href {https://doi.org/10.1103/physrevlett.91.180403} {\bibfield
  {journal} {\bibinfo  {journal} {Physical Review Letters}\ }\textbf {\bibinfo
  {volume} {91}},\ \bibinfo {pages} {180403} (\bibinfo {year}
  {2003})}\BibitemShut {NoStop}%
\bibitem [{\citenamefont {Gibilisco}\ and\ \citenamefont
  {Isola}(2006)}]{Gibilisco2006}%
  \BibitemOpen
  \bibfield  {author} {\bibinfo {author} {\bibfnamefont {P.}~\bibnamefont
  {Gibilisco}}\ and\ \bibinfo {author} {\bibfnamefont {T.}~\bibnamefont
  {Isola}},\ }\href {https://doi.org/10.1007/s10463-006-0103-3} {\bibfield
  {journal} {\bibinfo  {journal} {Annals of the Institute of Statistical
  Mathematics}\ }\textbf {\bibinfo {volume} {59}},\ \bibinfo {pages} {147}
  (\bibinfo {year} {2006})}\BibitemShut {NoStop}%
\bibitem [{\citenamefont {Gibilisco}\ \emph {et~al.}(2007)\citenamefont
  {Gibilisco}, \citenamefont {Imparato},\ and\ \citenamefont
  {Isola}}]{Gibilisco2007}%
  \BibitemOpen
  \bibfield  {author} {\bibinfo {author} {\bibfnamefont {P.}~\bibnamefont
  {Gibilisco}}, \bibinfo {author} {\bibfnamefont {D.}~\bibnamefont
  {Imparato}},\ and\ \bibinfo {author} {\bibfnamefont {T.}~\bibnamefont
  {Isola}},\ }\bibfield  {journal} {\bibinfo  {journal} {Journal of
  Mathematical Physics}\ }\textbf {\bibinfo {volume} {48}},\ \href
  {https://doi.org/10.1063/1.2748210} {10.1063/1.2748210} (\bibinfo {year}
  {2007})\BibitemShut {NoStop}%
\bibitem [{\citenamefont {Tonchev}(2021)}]{Tonchev2021}%
  \BibitemOpen
  \bibfield  {author} {\bibinfo {author} {\bibfnamefont {N.~S.}\ \bibnamefont
  {Tonchev}},\ }\bibfield  {journal} {\bibinfo  {journal} {arXiv preprint
  arXiv:2106.07599}\ }\href {https://doi.org/10.48550/arXiv.2106.07599}
  {10.48550/arXiv.2106.07599} (\bibinfo {year} {2021})\BibitemShut {NoStop}%
\bibitem [{Note5()}]{Note5}%
  \BibitemOpen
  \bibinfo {note} {This can be verified directly from the Lehmann
  representation $S(\omega )=\DOTSB \sum@ \slimits@ _{E_m\protect \neq E_n}
  (p_m+p_n) |O_{mn}-\delta _{mn}\mathinner {\langle {\protect \hat {O}}\rangle
  }|^2 \pi \delta (\omega +E_m-E_n)$}\BibitemShut {NoStop}%
\bibitem [{\citenamefont {Holevo}(1973)}]{Holevo73}%
  \BibitemOpen
  \bibfield  {author} {\bibinfo {author} {\bibfnamefont {A.~S.}\ \bibnamefont
  {Holevo}},\ }\href {https://doi.org/10.1137/1118039} {\bibfield  {journal}
  {\bibinfo  {journal} {Theory of Probability and Its Applications}\ }\textbf
  {\bibinfo {volume} {18}},\ \bibinfo {pages} {359} (\bibinfo {year}
  {1973})}\BibitemShut {NoStop}%
\bibitem [{\citenamefont {Kholevo}(1974)}]{kholevo1974}%
  \BibitemOpen
  \bibfield  {author} {\bibinfo {author} {\bibfnamefont {A.}~\bibnamefont
  {Kholevo}},\ }\href@noop {} {\bibfield  {journal} {\bibinfo  {journal}
  {Theory of Probability \& Its Applications}\ }\textbf {\bibinfo {volume}
  {18}},\ \bibinfo {pages} {359} (\bibinfo {year} {1974})},\ \bibinfo {note}
  {translated from the Russian original, published in 1973}\BibitemShut
  {NoStop}%
\bibitem [{\citenamefont {SCHULTZ}\ \emph {et~al.}(1964)\citenamefont
  {SCHULTZ}, \citenamefont {MATTIS},\ and\ \citenamefont {LIEB}}]{SCHULTZ1964}%
  \BibitemOpen
  \bibfield  {author} {\bibinfo {author} {\bibfnamefont {T.~D.}\ \bibnamefont
  {SCHULTZ}}, \bibinfo {author} {\bibfnamefont {D.~C.}\ \bibnamefont
  {MATTIS}},\ and\ \bibinfo {author} {\bibfnamefont {E.~H.}\ \bibnamefont
  {LIEB}},\ }\href {https://doi.org/10.1103/revmodphys.36.856} {\bibfield
  {journal} {\bibinfo  {journal} {Reviews of Modern Physics}\ }\textbf
  {\bibinfo {volume} {36}},\ \bibinfo {pages} {856} (\bibinfo {year}
  {1964})}\BibitemShut {NoStop}%
\bibitem [{\citenamefont {Derzhko}\ and\ \citenamefont
  {Krokhmalskii}(1997)}]{Derzhko1997}%
  \BibitemOpen
  \bibfield  {author} {\bibinfo {author} {\bibfnamefont {O.}~\bibnamefont
  {Derzhko}}\ and\ \bibinfo {author} {\bibfnamefont {T.}~\bibnamefont
  {Krokhmalskii}},\ }\href {https://doi.org/10.1103/physrevb.56.11659}
  {\bibfield  {journal} {\bibinfo  {journal} {Physical Review B}\ }\textbf
  {\bibinfo {volume} {56}},\ \bibinfo {pages} {11659} (\bibinfo {year}
  {1997})}\BibitemShut {NoStop}%
\bibitem [{\citenamefont {Sachdev}(2011)}]{Sachdev2011}%
  \BibitemOpen
  \bibfield  {author} {\bibinfo {author} {\bibfnamefont {S.}~\bibnamefont
  {Sachdev}},\ }\href {https://doi.org/10.1017/cbo9780511973765} {\emph
  {\bibinfo {title} {Quantum Phase Transitions}}}\ (\bibinfo  {publisher}
  {Cambridge University Press},\ \bibinfo {year} {2011})\BibitemShut {NoStop}%
\bibitem [{\citenamefont {Braunstein}\ and\ \citenamefont
  {Caves}(1994)}]{Braunstein1994}%
  \BibitemOpen
  \bibfield  {author} {\bibinfo {author} {\bibfnamefont {S.~L.}\ \bibnamefont
  {Braunstein}}\ and\ \bibinfo {author} {\bibfnamefont {C.~M.}\ \bibnamefont
  {Caves}},\ }\href {https://doi.org/10.1103/physrevlett.72.3439} {\bibfield
  {journal} {\bibinfo  {journal} {Physical Review Letters}\ }\textbf {\bibinfo
  {volume} {72}},\ \bibinfo {pages} {3439} (\bibinfo {year}
  {1994})}\BibitemShut {NoStop}%
\bibitem [{\citenamefont {Lieb}\ and\ \citenamefont
  {Robinson}(1972)}]{Lieb1972}%
  \BibitemOpen
  \bibfield  {author} {\bibinfo {author} {\bibfnamefont {E.~H.}\ \bibnamefont
  {Lieb}}\ and\ \bibinfo {author} {\bibfnamefont {D.~W.}\ \bibnamefont
  {Robinson}},\ }\href {https://doi.org/10.1007/bf01645779} {\bibfield
  {journal} {\bibinfo  {journal} {Communications in Mathematical Physics}\
  }\textbf {\bibinfo {volume} {28}},\ \bibinfo {pages} {251} (\bibinfo {year}
  {1972})}\BibitemShut {NoStop}%
\bibitem [{\citenamefont {Nachtergaele}\ and\ \citenamefont
  {Sims}(2006)}]{Nachtergaele2006}%
  \BibitemOpen
  \bibfield  {author} {\bibinfo {author} {\bibfnamefont {B.}~\bibnamefont
  {Nachtergaele}}\ and\ \bibinfo {author} {\bibfnamefont {R.}~\bibnamefont
  {Sims}},\ }\href {https://doi.org/10.1007/s00220-006-1556-1} {\bibfield
  {journal} {\bibinfo  {journal} {Communications in Mathematical Physics}\
  }\textbf {\bibinfo {volume} {265}},\ \bibinfo {pages} {119} (\bibinfo {year}
  {2006})}\BibitemShut {NoStop}%
\bibitem [{\citenamefont {Louisell}(1963)}]{Louisell1963}%
  \BibitemOpen
  \bibfield  {author} {\bibinfo {author} {\bibfnamefont {W.}~\bibnamefont
  {Louisell}},\ }\href {https://doi.org/10.1016/0031-9163(63)90442-6}
  {\bibfield  {journal} {\bibinfo  {journal} {Physics Letters}\ }\textbf
  {\bibinfo {volume} {7}},\ \bibinfo {pages} {60} (\bibinfo {year}
  {1963})}\BibitemShut {NoStop}%
\bibitem [{\citenamefont {Carruthers}\ and\ \citenamefont
  {Nieto}(1968)}]{Carruthers1968}%
  \BibitemOpen
  \bibfield  {author} {\bibinfo {author} {\bibfnamefont {P.}~\bibnamefont
  {Carruthers}}\ and\ \bibinfo {author} {\bibfnamefont {M.~M.}\ \bibnamefont
  {Nieto}},\ }\href {https://doi.org/10.1103/revmodphys.40.411} {\bibfield
  {journal} {\bibinfo  {journal} {Reviews of Modern Physics}\ }\textbf
  {\bibinfo {volume} {40}},\ \bibinfo {pages} {411} (\bibinfo {year}
  {1968})}\BibitemShut {NoStop}%
\bibitem [{\citenamefont {Callen}\ and\ \citenamefont
  {Welton}(1951)}]{Callen1951}%
  \BibitemOpen
  \bibfield  {author} {\bibinfo {author} {\bibfnamefont {H.~B.}\ \bibnamefont
  {Callen}}\ and\ \bibinfo {author} {\bibfnamefont {T.~A.}\ \bibnamefont
  {Welton}},\ }\href {https://doi.org/10.1103/physrev.83.34} {\bibfield
  {journal} {\bibinfo  {journal} {Physical Review}\ }\textbf {\bibinfo {volume}
  {83}},\ \bibinfo {pages} {34} (\bibinfo {year} {1951})}\BibitemShut {NoStop}%
\bibitem [{\citenamefont {Kato}(1995)}]{Kato1995}%
  \BibitemOpen
  \bibfield  {author} {\bibinfo {author} {\bibfnamefont {T.}~\bibnamefont
  {Kato}},\ }\href {https://doi.org/10.1007/978-3-642-66282-9} {\emph {\bibinfo
  {title} {Perturbation Theory for Linear Operators}}}\ (\bibinfo  {publisher}
  {Springer Berlin Heidelberg},\ \bibinfo {year} {1995})\BibitemShut {NoStop}%
\bibitem [{\citenamefont {Lancaster}\ and\ \citenamefont
  {Tismenetsky}(2007)}]{Lancaster2007}%
  \BibitemOpen
  \bibfield  {author} {\bibinfo {author} {\bibfnamefont {P.}~\bibnamefont
  {Lancaster}}\ and\ \bibinfo {author} {\bibfnamefont {M.}~\bibnamefont
  {Tismenetsky}},\ }\href@noop {} {\emph {\bibinfo {title} {The theory of
  matrices}}},\ \bibinfo {edition} {2nd}\ ed.,\ Computer science and applied
  mathematics\ (\bibinfo  {publisher} {Academic Press},\ \bibinfo {address}
  {San Diego, Fla. [u.a.]},\ \bibinfo {year} {2007})\BibitemShut {NoStop}%
\bibitem [{\citenamefont {Altland}\ and\ \citenamefont
  {Simons}(2010)}]{Altland2010}%
  \BibitemOpen
  \bibfield  {author} {\bibinfo {author} {\bibfnamefont {A.}~\bibnamefont
  {Altland}}\ and\ \bibinfo {author} {\bibfnamefont {B.~D.}\ \bibnamefont
  {Simons}},\ }\href {https://doi.org/10.1017/cbo9780511789984} {\emph
  {\bibinfo {title} {Condensed Matter Field Theory}}}\ (\bibinfo  {publisher}
  {Cambridge University Press},\ \bibinfo {year} {2010})\BibitemShut {NoStop}%
\bibitem [{Note6()}]{Note6}%
  \BibitemOpen
  \bibinfo {note} {We use $e^{-\mu |t|}$ kernel for simplicity. The physical
  kernel $g_\beta (t)$ (from Eq.~\protect \textup {\hbox {\mathsurround \z@
  \protect \normalfont (\ignorespaces \ref {eq:L_as_dressed_O}\unskip
  \@@italiccorr )}}) has the same large-$t$ exponential decay ($\sim e^{-(\pi
  /\beta )|t|}$), so replacing $g_\beta (t)$ with $e^{-\mu |t|}$ (with $\mu =
  \pi /\beta $) does not affect the final exponential nature of the
  bound.}\BibitemShut {Stop}%
\bibitem [{\citenamefont {Bachmann}\ \emph {et~al.}(2011)\citenamefont
  {Bachmann}, \citenamefont {Michalakis}, \citenamefont {Nachtergaele},\ and\
  \citenamefont {Sims}}]{Bachmann2011}%
  \BibitemOpen
  \bibfield  {author} {\bibinfo {author} {\bibfnamefont {S.}~\bibnamefont
  {Bachmann}}, \bibinfo {author} {\bibfnamefont {S.}~\bibnamefont
  {Michalakis}}, \bibinfo {author} {\bibfnamefont {B.}~\bibnamefont
  {Nachtergaele}},\ and\ \bibinfo {author} {\bibfnamefont {R.}~\bibnamefont
  {Sims}},\ }\href {https://doi.org/10.1007/s00220-011-1380-0} {\bibfield
  {journal} {\bibinfo  {journal} {Communications in Mathematical Physics}\
  }\textbf {\bibinfo {volume} {309}},\ \bibinfo {pages} {835} (\bibinfo {year}
  {2011})}\BibitemShut {NoStop}%
\bibitem [{\citenamefont {Nachtergaele}\ \emph {et~al.}(2012)\citenamefont
  {Nachtergaele}, \citenamefont {Scholz},\ and\ \citenamefont
  {Werner}}]{Nachtergaele2012}%
  \BibitemOpen
  \bibfield  {author} {\bibinfo {author} {\bibfnamefont {B.}~\bibnamefont
  {Nachtergaele}}, \bibinfo {author} {\bibfnamefont {V.~B.}\ \bibnamefont
  {Scholz}},\ and\ \bibinfo {author} {\bibfnamefont {R.~F.}\ \bibnamefont
  {Werner}},\ }\bibinfo {title} {Local approximation of observables and
  commutator bounds},\ in\ \href {https://doi.org/10.1007/978-3-0348-0531-5_8}
  {\emph {\bibinfo {booktitle} {Operator Methods in Mathematical Physics}}}\
  (\bibinfo  {publisher} {Springer Basel},\ \bibinfo {year} {2012})\ pp.\
  \bibinfo {pages} {143--149}\BibitemShut {NoStop}%
\bibitem [{Note7()}]{Note7}%
  \BibitemOpen
  \bibinfo {note} {For the operator $\protect \hat {L}$ defined by Eq.~\protect
  \textup {\hbox {\mathsurround \z@ \protect \normalfont (\ignorespaces \ref
  {eq:L_as_dressed_O}\unskip \@@italiccorr )}}, its boundedness is
  correspondingly ensured by the integral $ I = \DOTSI \intop \ilimits@
  _{-\infty }^{+\infty } dt \protect \, g_{\beta }(t)= \protect \frac {4}{\pi }
  \DOTSI \intop \ilimits@ _{0}^{\infty } \protect \qopname \relax o{ln}\left
  [\protect \qopname \relax o{tanh}\left (\protect \frac {\pi t}{2\beta }\right
  )\right ] dt $. The substitution $u = \protect \qopname \relax o{tanh}\left
  (\protect \frac {\pi t}{2\beta }\right )$ yields a finite result $I =
  \protect \frac {8\beta }{\pi ^2} \DOTSI \intop \ilimits@ _0^1 \protect \frac
  {\protect \qopname \relax o{ln}u}{1-u^2} du = -\beta $ which confirms the
  boundedness of $\protect \hat {L}$.}\BibitemShut {Stop}%
\bibitem [{Note8()}]{Note8}%
  \BibitemOpen
  \bibinfo {note} {For an infinite lattice, $\protect \hat {L}$ is an extensive
  quantity. For such operators, the relevant metric is not the total error but
  the error per site. Our construction, which replaces each $\protect \hat
  {L}_{i}$ with its local approximation, provides a concrete scheme that
  guarantees the error per site is exponentially small.}\BibitemShut {Stop}%
\end{thebibliography}%

\clearpage
\onecolumngrid
\beginsupplement

\section*{Supplemental Material for ``Bounds on quantum Fisher information and uncertainty relations for thermodynamically conjugate variables''}

\section{Quantum Fisher information for Gibbs ensembles} \label{sec:QFI_Gibbs}
This section details the derivation of the quantum Fisher Information for a Gibbs state,
expressing it via its conjugate observable, as defined by Eq.~\eqref{eq:QFI1} in the main text.
The system is described by the Gibbs density matrix
\begin{equation}
    \hat{\rho}_\theta = \frac{e^{-\beta \hat{H}(\theta)}}{\text{Tr}[e^{-\beta \hat{H}(\theta)}]},
\end{equation}
where the parameter $\theta$ is encoded in the Hamiltonian $\hat{H}(\theta)$.
The starting point for our analysis is the standard spectral representation of the quantum Fisher information for the state $\hat{\rho}_\theta$:
\begin{equation}\label{eq:QFI2}
\mathcal{F}_\theta = \sum_n \frac{(\partial_\theta p_n)^2}{p_n}+\sum_{m \neq n} \frac{2(p_m-p_n)^2}{p_m+p_n} \vert \braket{m|\partial_\theta n} \vert^2,
\end{equation}
where $\ket{n}$ and $E_n$ are the eigenstates and energy eigenvalues of the Hamiltonian $\hat{H}(\theta)$,
respectively, defined by the eigenvalue equation $\hat{H} \ket{n} = E_n \ket{n}$.
The terms $p_n = e^{-\beta E_n}/\sum_ne^{-\beta E_n}$ are the corresponding Gibbs populations.
Our goal is to rewrite this quantum Fisher information in terms of the matrix elements of the conjugate observable, $\hat{O} = \partial_\theta \hat{H}(\theta)$,
thereby eliminating the explicit partial derivatives.

Let us first consider the first term of Eq.~\eqref{eq:QFI2}.
The derivative of the Gibbs populations $\partial_\theta p_n$ is given by
\begin{align}
    \partial_\theta p_n=p_n \partial_\theta \ln p_n=p_n\partial_\theta (-\beta E_n-\ln (\sum_k e^{-\beta E_k}))=-\beta p_n(O_{nn}-\braket{\hat{O}}), \label{eq:qfi-simplify-term1}
\end{align}
where $ O_{nn}=\braket{n|\hat{O}|n}=\partial_\theta E_n$ and $\braket{\hat{O}}=\sum_np_n O_{nn}$.
Substituting this result back into the first term of Eq.~\eqref{eq:QFI2} yields
\begin{equation}
    \sum_n \frac{(\partial_\theta p_n)^2}{p_n} = \beta^2 \sum_n p_n(O_{nn}-\braket{\hat{O}})^2. \label{eq:term1}
\end{equation}

In evaluating the second term of Eq.~\eqref{eq:QFI2}, we first establish a general relation for the matrix elements $\braket{m|\partial_\theta n}$. We begin by differentiating the eigenvalue equation $\hat{H} \ket{n} = E_n \ket{n}$ with respect to $\theta$:
\begin{equation}
    (\partial_\theta \hat{H})\ket{n} + \hat{H}(\partial_\theta \ket{n}) = (\partial_\theta E_n)\ket{n} + E_n (\partial_\theta \ket{n}).
\end{equation}
Taking the inner product with $\bra{m}$ for $m \neq n$ yields
\begin{align}
    \braket{m|\hat{O}|n} + E_m\braket{m|\partial_\theta n} = E_n \braket{m|\partial_\theta n},
\end{align}
which can be rearranged into the central relation
\begin{equation} \label{eq:Omn_relation}
    O_{mn} = (E_n - E_m)\braket{m|\partial_\theta n}.
\end{equation}
For non-degenerate states where $E_m \neq E_n$,
relation Eq.~\eqref{eq:Omn_relation} leads to the well-known first-order perturbation theory:
\begin{equation}
    \braket{m|\partial_\theta n}=\frac{O_{mn}}{E_n-E_m}. \label{eq:qfi-simplify-term2}
\end{equation}
For degenerate states where $E_m = E_n$,
the right-hand side of Eq.~\eqref{eq:Omn_relation} vanishes.
As one can always choose a proper gauge to ensure that the derivative term $\braket{m|\partial_\theta n}$ is finite~\citep{Kato1995, Lancaster2007},
Eq.~\eqref{eq:Omn_relation} immediately requires that
\begin{equation}
    O_{mn} = 0 \quad (\text{for } E_m = E_n, m \neq n). \label{eq:Omn_zero}
\end{equation}
This result is consistent with degenerate perturbation theory,
which requires that the perturbation operator $\hat{O}$ be diagonal within the basis of the degenerate subspace.

The analysis for the non-degenerate case [Eq.~\eqref{eq:qfi-simplify-term2}] rewrite the second term of Eq.~\eqref{eq:QFI2} as follows:
\begin{equation} \label{eq:term2}
    \sum_{m \neq n} \frac{2(p_m-p_n)^2}{p_m+p_n} \vert \braket{m|\partial_\theta n} \vert^2= 
    2 \sum_{E_m \neq E_n}\frac{(p_m-p_n)^2}{p_m+p_n} \frac{1}{(E_m-E_n)^2} |O_{mn}|^2,
\end{equation}
Taken together, Eq.~\eqref{eq:term1} and Eq.~\eqref{eq:term2} gives the final expression for the quantum Fisher information
\begin{equation}\label{eq:QFI3}
    \mathcal{F}_\theta = \beta^2 \sum_n p_n(O_{nn}-\braket{\hat{O}})^2 + 2 \sum_{E_m \neq E_n}\frac{(p_m-p_n)^2}{p_m+p_n} \frac{1}{(E_m-E_n)^2} |O_{mn}|^2.
\end{equation}

\section{Fluctuation-dissipation relation} \label{sec:FDT_generalized}
In this section, we derive the generalized fluctuation-dissipation theorem that connects the two integral representations for the quantum Fisher information from the main text:
the form involving the Kubo response [Eq.~\eqref{eq:QFI_dissipation}] and the one involving the autocorrelation spectrum [Eq.~\eqref{eq:QFI_fluctuation}].
However, a naive application of the standard Callen-Welton fluctuation-dissipation relation, $S(\omega)=\coth(\beta\omega/2) \text{Im}[\chi(\omega)]$,
is insufficient due to subtleties arising at zero frequency that relate to the first term in Eq.~\eqref{eq:QFI_dissipation}.

To derive the correct, generalized fluctuation-dissipation relation, we start with the definition of the time-ordered Green's function $C^T(\omega)$, the retarded Green's function $C^R(\omega)$, and the symmetrized autocorrelation spectrum $S(\omega)$ as follows,
\begin{align}
    C^T(\omega)&=-i \int_{-\infty}^{+\infty} dt e^{i\omega t} \braket{\mathcal{T} \Delta \hat{O}(t) \Delta \hat{O}} \label{eq:defct} \\
    C^R(\omega)&=-i \int_{-\infty}^{+\infty} dt e^{i\omega t} \Theta(t) \braket{[\hat{O}(t), \hat{O}]} \label{eq:defcr} \\
    S(\omega)  &=\frac{1}{2}\int_{-\infty}^{+\infty} dt e^{i\omega t} \braket{\Delta \hat{O}(t) \Delta \hat{O}}+\braket{\Delta \hat{O} \Delta \hat{O} (t)}.
\end{align}
The autocorrelation function $S(\omega)$ can link to the imaginary part of the time-ordered Green's function,
\begin{align}
S(\omega)=&\frac{1}{2}\int_{-\infty}^{+\infty} dt e^{i\omega t} \braket{\Delta \hat{O}(t) \Delta \hat{O}}+\braket{\Delta \hat{O} \Delta \hat{O} (t)} \nonumber \\
=&\frac{1}{2}\left[
\int_{-\infty}^{0} dt e^{i\omega t} \braket{\Delta \hat{O}(t) \Delta \hat{O}} + \int_{0}^{+\infty} dt e^{i\omega t} \braket{\Delta \hat{O}(t) \Delta \hat{O}}
+\int_{-\infty}^{0} dt e^{i\omega t} \braket{\Delta \hat{O} \Delta \hat{O} (t)} +\int_{0}^{+\infty} dt e^{i\omega t} \braket{\Delta \hat{O} \Delta \hat{O} (t)}
\right] \nonumber \\
=&\frac{1}{2}\left[
 \int_{0}^{+\infty} dt e^{-i\omega t} \braket{\Delta \hat{O} \Delta \hat{O}(t)} + \int_{0}^{+\infty} dt e^{i\omega t} \braket{\Delta \hat{O}(t) \Delta \hat{O}}
+\int_{-\infty}^{0} dt e^{i\omega t} \braket{\Delta \hat{O} \Delta \hat{O} (t)} + \int_{-\infty}^{0} dt e^{-i\omega t} \braket{\Delta \hat{O}(t) \Delta \hat{O}}
\right] \nonumber \\
=&\frac{1}{2}\left[
(\int_{0}^{+\infty} dt e^{i\omega t} \braket{\Delta \hat{O}(t) \Delta \hat{O}}
+\int_{-\infty}^{0} dt e^{i\omega t} \braket{\Delta \hat{O} \Delta \hat{O} (t)})
+
(\int_{0}^{+\infty} dt e^{i\omega t} \braket{\Delta \hat{O}(t) \Delta \hat{O}}
+\int_{-\infty}^{0} dt e^{i\omega t} \braket{\Delta \hat{O} \Delta \hat{O} (t)})^*
\right] \nonumber \\
=&-\frac{C^T(\omega)-C^T(\omega)^*}{2i} \nonumber \\
=&-\text{Im}[C^T(\omega)].\label{eq:imagct}
\end{align}

The dissipative response function, $\chi''(\omega)$, linking to the imaginary part of the retarded Green's function as a consequence of linear response theory~\citep{Altland2010, Coleman2015}
\begin{equation}
\text{Im}[\chi(\omega)] = -\text{Im}[C^R(\omega)].\label{eq:imagcr}
\end{equation}
The precise relationship between $C^T(\omega)$ and $C^R(\omega)$ is revealed in the Lehmann representation of Eq.~\eqref{eq:defct} and Eq.~\eqref{eq:defcr},
\begin{align}
C^T(\omega)=&\lim_{\eta\to 0^+} \sum_{mn} |O_{mn}-\delta_{mn}\braket{\hat{O}}|^2 \left[ \frac{p_m}{\omega+E_m-E_n+i\eta} -\frac{p_n}{\omega+E_m-E_n-i\eta} \right] \\
C^R(\omega)=&\lim_{\eta\to 0^+} \sum_{mn} |O_{mn}-\delta_{mn}\braket{\hat{O}}|^2 \left[ \frac{p_m}{\omega+E_m-E_n+i\eta} -\frac{p_n}{\omega+E_m-E_n+i\eta} \right].
\end{align}
Note the retarded expression also contains the $\delta_{mn}\braket{\hat{O}}$ term,
this is because there is no difference between the retarded Green's function between $O$ and $\Delta O=O-\braket{\hat{O}}$,
the $\delta_{mn}\braket{\hat{O}}$ term only affects $m=n$ terms in summation, but these terms are always zero since $p_m=p_n$ for $m=n$.
Taking the imaginary part while combining with Eq.~\eqref{eq:imagct} and Eq.~\eqref{eq:imagcr},
\begin{align}
S(\omega)=-\text{Im}[C^T(\omega)]=& \sum_{mn} (p_m+p_n) |O_{mn}-\delta_{mn}\braket{\hat{O}}|^2 \pi \delta(\omega+E_m-E_n) \label{eq:lehmannimct} \\
\text{Im}[\chi(\omega)]=-\text{Im}[C^R(\omega)]=& \sum_{mn} (p_m-p_n) |O_{mn}-\delta_{mn}\braket{\hat{O}}|^2 \pi \delta(\omega+E_m-E_n) \label{eq:lehmannimcr}.
\end{align}

For any non-zero frequency ($\omega \neq 0$), the delta function fixes $E_n-E_m = \omega$. We can therefore relate the population factors,
\begin{align}
    (p_m+p_n)\delta(\omega+E_m-E_n)&=(p_m-p_n)\frac{p_m+p_n}{p_m-p_n}\delta(\omega+E_m-E_n) \nonumber \\
    &=(p_m-p_n)\coth(\frac{\beta(E_n-E_m)}{2})\delta(\omega+E_m-E_n) \nonumber \\
    &=(p_m-p_n)\coth(\frac{\beta \omega}{2})\delta(\omega+E_m-E_n).
\end{align}
which leads to the standard fluctuation-dissipation relation,
\begin{equation}
    S(\omega)=\coth(\frac{\beta\omega}{2})\text{Im}[\chi(\omega)] \text{,\;for $\omega \neq 0$} \label{eq:fdt_neq0}.
\end{equation}
At exactly zero frequency, the two factors appearing in Eq.~\eqref{eq:lehmannimcr} --- the Dirac delta $\delta(\omega+E_m-E_n)$ and the population difference $(p_m-p_n)$ --- cannot be simultaneously nonzero:
$\delta(0+E_m-E_n) \neq 0$ enforces $E_m=E_n$, for which $p_m-p_n=0$; conversely, whenever $p_m\neq p_n$ one must have $E_m\neq E_n$ and hence $\delta(0+E_m-E_n)=0$.
As a consequence, the imaginary part of the Kubo response has no zero-frequency singularity and $\lim_{\omega\to 0}\text{Im}[\chi(\omega)]=0$.
In contrast, the autocorrelation spectrum has a well-defined zero-frequency component obtained by isolating the contributions with $E_m=E_n$:
\begin{align}
    \lim_{\omega\to 0}S(\omega)=&\pi \delta(\omega) \sum_{E_m=E_n} (p_m+p_n) |O_{mn}-\delta_{mn}\braket{\hat{O}}|^2 \nonumber \\
    =&2\pi \delta(\omega) \sum_n p_n (O_{nn}-\braket{\hat{O}})^2.
\end{align}
The second equality holds because the off-diagonal terms $O_{mn}$
vanish for degenerate states $(E_m=E_n)$ when $m\neq n$
as established in Sec.~\ref{sec:QFI_Gibbs}.

The generalized fluctuation-dissipation relation is therefore obtained by augmenting the standard relation for $\omega \neq 0$ with this singular, zero-frequency term:
\begin{equation}
    S(\omega)=\coth(\frac{\beta\omega}{2})\text{Im}[\chi(\omega)] + 2\pi\delta(\omega) \sum_{n}p_n (O_{nn}-\braket{\hat{O}})^2.
\end{equation}
Substituting this generalized fluctuation-dissipation relation into Eq.\eqref{eq:QFI_dissipation}
directly yields Eq.~\eqref{eq:QFI_fluctuation} in the main text:
\begin{align}
\mathcal{F}_\theta=& \beta^2 \sum_n p_n(O_{nn}-\braket{\hat{O}})^2 + \frac{2}{\pi} \int_{-\infty}^{+\infty} d\omega \tanh(\frac{\omega \beta}{2}) \frac{1}{\omega^2} \text{Im}[\chi(\omega)].
\end{align}
We note that Ref.~\citep{Tonchev2021} also derives an integral representation for the quantum Fisher information in terms of the Kubo response function. However, because they employ the standard Callen-Welton relation --- without the zero-frequency correction --- their integral representation of the Kubo response function does not capture the first, purely statistical term present in Eq.~\eqref{eq:QFI_dissipation}.

\section{The symmetric logarithmic derivative operator for Gibbs ensembles} \label{sec:SLD}
Here, we provide a detailed derivation for the integral representation of the symmetric logarithmic
derivative operator of a Gibbs state, presented as Eq.~\eqref{eq:SLD_integral} in the main text.
We begin with the SLD operator $\hat{L}$ defined by the Lyapunov equation
\begin{align}\label{eq:SLD}
    \frac{1}{2}(\hat{\rho}_\theta \hat{L}+\hat{L} \hat{\rho}_\theta)
    = \partial_\theta \hat{\rho}_\theta.
\end{align}
The derivative of the density matrix $\hat{\rho}_\theta$ with respect to a parameter $\theta$ is given by the integral representation
\begin{align}
    \partial_\theta \hat{\rho}_\theta
    = - \beta \int_0^1 d\lambda \, \hat{\rho}_\theta^\lambda (\hat{O} - \braket{ \hat{O} })
    \hat{\rho}_\theta^{1-\lambda}.
\end{align}
Without loss of generality, we set $\braket{\hat{O}} = \text{Tr}[\hat{\rho}_\theta \hat{O}] = 0$.
This is justified because any component of $\hat{O}$ proportional to the identity operator does
not contribute to the derivative, as $\partial_\theta \hat{\rho}_\theta$ is traceless.
The integral representation thus simplifies to
$\partial_\theta \hat{\rho}_\theta = - \beta \int_0^1 d\lambda \, \hat{\rho}_\theta^\lambda \hat{O} \hat{\rho}_\theta^{1-\lambda}$.
Substituting this into Eq.~\eqref{eq:SLD} and applying the change of variables $\tau=\beta \lambda$ yields:
\begin{align}\label{eq:SLD_O_integral}
    \frac{1}{2}(\hat{\rho}_\theta \hat{L}+\hat{L} \hat{\rho}_\theta)
    = -\frac{1}{Z} \int_0^\beta d\tau \, e^{-\tau \hat{H}(\theta)} \hat{O} e^{-(\beta-\tau) \hat{H}(\theta)},
\end{align}
where $Z(\theta)=\text{Tr}[e^{-\beta \hat{H}(\theta)}]$ is the partition function.
For simplicity, we omit the $\theta$ dependence of the Hamiltonian and the partition function using the notation $\hat{H}$ and $Z$ for the remainder of the derivation.

We now solve this equation for the matrix elements $L_{mn} \equiv \braket{ m|\hat{L}|n }$ in the energy eigenbasis $\{\ket{n}\}$.
For the non-degenerate elements ($E_m \neq E_n$), taking the matrix elements of Eq.~\eqref{eq:SLD_O_integral} yields
\begin{align}
        \frac{1}{2}(p_m+p_n)L_{mn}
        &= -\frac{1}{Z} \int_0^\beta d\tau \, \braket{ m| e^{-\tau \hat{H}} \hat{O} e^{-(\beta-\tau) \hat{H}} |n } \nonumber \\
        &= -\frac{e^{-\beta E_n}}{Z} \int_0^\beta d\tau \, O_{mn} e^{-\tau (E_m-E_n)} \nonumber \\
        &= -\frac{1}{Z} \frac{e^{-\beta E_m}-e^{-\beta E_n}}{E_n-E_m}O_{mn} \nonumber \\
        &= \frac{p_m-p_n}{E_m-E_n} O_{mn}. \label{eq:SLD_mat_offdiag_pre}
\end{align}
We define the energy-domain weighting kernel $f(\omega)$ as
\begin{align}
    f(\omega) = -\frac{\tanh(\beta\omega/2)}{\omega/2},
\end{align}
which rewrites Eq.~\eqref{eq:SLD_mat_offdiag_pre} concisely as
\begin{align}\label{eq:SLD_mat_offdiag}
    L_{mn} &= \frac{2(p_m - p_n)}{(p_m+p_n)(E_m-E_n)} O_{mn} \nonumber \\
    &= f(E_m-E_n) O_{mn}
\end{align}
For the degenerate elements ($E_m=E_n$),
it follows that $p_m=p_n$,
and the matrix elements of Eq.~\eqref{eq:SLD_O_integral} become
\begin{align}
    p_n L_{mn} &=
    -\frac{1}{Z}
    \int_0^\beta d\tau \, \braket{ m| e^{-\tau \hat{H}} \hat{O} e^{-(\beta-\tau) \hat{H}} |n } \nonumber \\
    &=-\beta p_n O_{mn}.
\end{align}
This yields the result
    \begin{align}\label{eq:SLD_mat_diag}
    L_{mn}=-\beta O_{mn}.
\end{align}
We now consider the zero-frequency limit of the kernel,
\begin{align}
    \lim_{\omega\to 0}f(\omega) = -\beta.
\end{align}
This result allows the degenerate [Eq.~\eqref{eq:SLD_mat_diag}] and non-degenerate [Eq.~\eqref{eq:SLD_mat_offdiag}] terms to be written in a single unified expression,
\begin{align}
    L_{mn} = f(E_m-E_n) O_{mn},
\end{align}
which holds for both zero and non-zero values of $E_m-E_n$.

To obtain a time-domain representation, we introduce the function $g_\beta(t)$ as the inverse Fourier transform of the weighting kernel
\begin{align}\label{eq:g_beta_t}
    g_\beta(t) &= \frac{1}{2\pi} \int_{-\infty}^{+\infty} d\omega \, e^{-i\omega t} f(\omega) \nonumber \\
    &= \frac{2}{\pi} \ln\left[\tanh\left(\frac{\pi|t|}{2\beta}\right)\right].
\end{align}
With this definition,
we rewrite the matrix elements $L_{mn}$ using the forward Fourier transformation, $f(\omega) = \int_{-\infty}^{+\infty} dt \, g_\beta(t) e^{i\omega t}$, as
\begin{align}
L_{mn} &= f(E_m-E_n) O_{mn} \nonumber \\
&= \left[ \int_{-\infty}^{+\infty} dt \, g_\beta(t) e^{i(E_m-E_n)t} \right] O_{mn} \nonumber \\
&= \int_{-\infty}^{+\infty} dt \, g_\beta(t) \braket{ m| e^{i\hat{H}t} \hat{O} e^{-i\hat{H}t} |n }.
\end{align}
Since this relation holds for any pair of eigenstates, it implies the operator identity:
\begin{align}\label{eq:L_as_dressed_O}
    \hat{L}=\int_{-\infty}^{+\infty} dt \, g_\beta(t) \hat{O}(t),
\end{align}
where $\hat{O}(t)=e^{i\hat{H}t}\hat{O}e^{-i\hat{H}t}$ is the operator $\hat{O}$ in the Heisenberg picture.

We have derived a time-domain integral representation for the symmetric logarithmic derivative operator of a Gibbs state, given in Eq.~\eqref{eq:L_as_dressed_O} with the specific kernel $g_\beta(t)$. To the best of our knowledge, this representation is a new result. It serves as the primary tool in the following section for proving the locality of the optimal estimator.

\section{Locality of dressed local operators} \label{sec:locality}
In this section, we demonstrate that the locality of the symmetric logarithmic derivative operator
$\hat{L}$ (defined in Eq.~\eqref{eq:L_as_dressed_O})
is inherited from the original operator $\hat{O}$.
Specifically, we prove that if $\hat{O}$ is a sum of local operators,
then $\hat{L}$ can be written as a corresponding sum of operators, each satisfying an exponential decay bound on its commutator with distant operators.

To proceed formally, we first specify what we mean by a local operator.
An operator $\hat{O}_i$ is called local if its support, i.e., the set of sites on which it acts nontrivially, has a finite size that does not scale with the total system size.

For clarity, instead of analyzing the full sum $\hat{O}=\sum_i \hat{O}_i$,
we consider a single representative local term $\hat{O}_{\text{loc}}$ (one of the $\hat{O}_i$)
and establish the desired property for it.

\textbf{Proposition:}
Let $\hat{O}_{\text{loc}}$ be an operator with a finite support $X$,
and let $\hat{H}$ be a local Hamiltonian. For any $\mu > 0$, the corresponding ``dressed'' operator
\begin{align}\label{eq:L_integral_t}
\hat{L}_{\text{loc}} = \int_{-\infty}^{+\infty} dt \, e^{-\mu|t|} \hat{O}_{\text{loc}}(t),
\quad \text{where} \quad \hat{O}_{\text{loc}}(t) = e^{i\hat{H}t} \hat{O}_{\text{loc}} e^{-i\hat{H}t},
\end{align}
obeys an exponential decay bound on its commutator with distant operators. 
Specifically, there exist constants $C$ and $\lambda > 0$ such that
\begin{align}
    \| [\hat{L}_{\text{loc}}, \hat{B}] \| \le C \|\hat{O}_{\text{loc}}\| \|\hat{B}\| e^{-\lambda r}
\end{align}
for any operator $\hat{B}$ supported at a distance $r$ from $X$,
where $\| \cdot \|$ denotes the operator spectral norm, the decay rate $\lambda = \min(a, \mu/v)$ is determined by the Hamiltonian's Lieb-Robinson parameters ($a, v$) and the integral's decay factor $\mu$.

The operator $\hat{B}$ acts as a ``probe operator'' supported on a region distant from $X$.
The bound on the commutator $\|[\hat{L}_{\text{loc}}, \hat{B}]\|$ thus measures the extent to which $\hat{L}_{\text{loc}}$ acts non-trivially on distant regions.

\textbf{Proof:} The proof relies on the Lieb-Robinson bound~\cite{Lieb1972, Nachtergaele2006},
\begin{align}
\|[\hat{O}_{\text{loc}}(t), \hat{B}]\| \leq C_{LR} \|\hat{O}_{\text{loc}}\| \|\hat{B}\| e^{-a(r - v|t|)},
\end{align}
which constrains the spectral norm of the commutator of $\hat{O}_{\text{loc}}(t)$ with a distant probe operator $\hat{B}$.
Here, $C_{LR}$ is a constant, while $a$ and $v$ are the LR decay rate and velocity, respectively.

To prove the exponential decay for $\hat{L}_{\text{loc}}$,
we bound its commutator by applying the triangle inequality to its integral representation (Eq.~\eqref{eq:L_integral_t})~\footnote{
    We use $e^{-\mu|t|}$ kernel for simplicity.
    The physical kernel $g_\beta(t)$ (from Eq.~\eqref{eq:L_as_dressed_O}) has the same large-$t$ exponential decay ($\sim e^{-(\pi/\beta)|t|}$),
    so replacing $g_\beta(t)$ with $e^{-\mu|t|}$ (with $\mu = \pi/\beta$) does not affect the final exponential nature of the bound.
}:
\begin{align}
\|[\hat{L}_{\text{loc}}, \hat{B}]\| \le \int_{-\infty}^{\infty} dt \, e^{-\mu|t|} \, \|[\hat{O}_{\text{loc}}(t), \hat{B}]\|.
\end{align}
Splitting the integral at the characteristic time $t_c = r/v$,
which separates the integration domain into regions outside and inside the effective light cone.

For the region outside the light cone ($|t| < t_c$), the contribution $\mathcal{C}_{<}$ is bounded by
\begin{align}
\mathcal{C}_{<} \leq C_{LR} \|\hat{O}_{\text{loc}}\| \|\hat{B}\| e^{-ar} \int_{-t_c}^{t_c} dt \, e^{(av - \mu)|t|}.
\end{align}
Evaluation of the integral gives two cases:
\begin{itemize}
    \item For $av \neq \mu$, the bound is a sum of two exponentially decaying terms:
    \begin{align}
        \mathcal{C}_{<} \leq \frac{2 C_{LR}}{|av-\mu|} \|\hat{O}_{\text{loc}}\| \|\hat{B}\| \left| e^{-(\mu/v)r} - e^{-ar} \right|.
    \end{align}
    \item For $av = \mu$, the bound is also exponentially decaying:
    \begin{align}
        \mathcal{C}_{<} \leq \frac{2 C_{LR}}{v} \|\hat{O}_{\text{loc}}\| \|\hat{B}\| r e^{-ar}.
    \end{align}
\end{itemize}
In both scenarios, this contribution decays exponentially with the distance $r$.
The overall decay is governed by the slower of the two rates, i.e., by $e^{-\min(a, \mu/v)r}$.

For the region inside the light cone ($|t| \ge t_c$),
the Lieb-Robinson bound becomes trivial ($e^{-a(r-v|t|)} \ge 1$).
We therefore use the general bound for a commutator: $\|[\hat{A}, \hat{C}]\| \le 2\|\hat{A}\|\|\hat{C}\|$.
Since time evolution is unitary, $\|\hat{O}_{\text{loc}}(t)\| = \|\hat{O}_{\text{loc}}\|$.
The contribution $\mathcal{C}_{\ge}$ is thus bounded by

\begin{align}
\mathcal{C}_{\ge} &\leq \int_{|t| \ge t_c} dt \, e^{-\mu|t|} \left( 2 \|\hat{O}_{\text{loc}}\| \|\hat{B}\| \right) \nonumber \\
&= 4 \|\hat{O}_{\text{loc}}\| \|\hat{B}\| \int_{t_c}^{\infty} dt \, e^{-\mu t} \nonumber \\
&= \frac{4}{\mu} \|\hat{O}_{\text{loc}}\| \|\hat{B}\| e^{-\mu t_c}.
\end{align}

Substituting $t_c = r/v$, we find that this contribution also decays exponentially with distance,
\begin{align}
\mathcal{C}_{\ge} \le \frac{4}{\mu} \|\hat{O}_{\text{loc}}\| \|\hat{B}\| e^{-(\mu/v)r}.
\end{align}

Since both $\mathcal{C}_{<}$ and $\mathcal{C}_{\ge}$ decay exponentially with the distance $r$, their sum does as well,
\begin{align}\label{eq:commutator_bound}
\|[\hat{L}_{\text{loc}}, \hat{B}]\| \leq \mathcal{C}_{<} + \mathcal{C}_{\ge} \leq C \|\hat{O}_{\text{loc}}\| \|\hat{B}\| e^{-\min(a, \mu/v) \cdot r}.
\end{align}
By linearity, the dressed version of the full operator $\hat{O} = \sum_i \hat{O}_{i}$ is $\hat{L} = \sum_i \hat{L}_{i}$.
The proof above confirms that each term $\hat{L}_i$ has an exponentially decaying commutator bound, meaning the full operator $\hat{L}$ is a sum of terms with this property.

Having established that the commutators of $\hat{L}_{\text{loc}}$ decay exponentially with distance,
we now show that this property allows it to be well-approximated by an operator with strictly finite support.
The proof relies on a general theorem regarding local operator approximation, which is rigorously proven in works of Bachmann et al. and Nachtergaele et al.:

\textbf{Theorem (Local Operator Approximation)~\citep{Bachmann2011, Nachtergaele2012}:}
Let the Hilbert space of the lattice be decomposed with respect to a region $\Lambda_R$
as $\mathcal{H} = \mathcal{H}_{\Lambda_R} \otimes \mathcal{H}_{\Lambda_R^c}$.
If an operator $A \in \mathcal{B}(\mathcal{H})$ satisfies the bound $\|[A, \mathbb{I}_{\Lambda_R} \otimes B]\| \le \epsilon \|B\|$
for all bounded $B$ acting on $\mathcal{H}_{\Lambda_R^c}$,
then there exists an operator $A'$ supported entirely on $\Lambda_R$ such that 
\begin{align}
    \|A - A'\| \le 2\epsilon.
\end{align}
Here $\mathcal{B}(\mathcal{H})$ denotes the Banach space of bounded operators on $\mathcal{H}$.

To apply this theorem, the operator $\hat{L}_{\text{loc}}$ needs to be bounded to ensure it belongs to the Banach space $\mathcal{B}(\mathcal{H})$. This is shown by applying the triangle inequality for integrals to its definition in Eq.~\eqref{eq:L_integral_t}~\footnote{
    For the operator $\hat{L}$ defined by Eq.~\eqref{eq:L_as_dressed_O}, its boundedness is correspondingly ensured by the integral $ I = \int_{-\infty}^{+\infty} dt \, g_{\beta}(t)= \frac{4}{\pi} \int_{0}^{\infty} \ln\left[\tanh\left(\frac{\pi t}{2\beta}\right)\right] dt $. The substitution $u = \tanh\left(\frac{\pi t}{2\beta}\right)$ yields a finite result $I = \frac{8\beta}{\pi^2} \int_0^1 \frac{\ln u}{1-u^2} du = -\beta$ which confirms the boundedness of $\hat{L}$.
}:
\begin{align}
    \|\hat{L}_{\text{loc}}\| = \left\| \int_{-\infty}^{+\infty} dt \, e^{-\mu|t|} \hat{O}_{\text{loc}}(t) \right\| \le \int_{-\infty}^{+\infty} dt \, e^{-\mu|t|} \|\hat{O}_{\text{loc}}(t)\| = \frac{2}{\mu} \|\hat{O}_{\text{loc}}\|.
\end{align}

We define the approximation region $\Lambda_R$ based on the support of the initial operator $\hat{O}_{\text{loc}}$.
We consider that the support of $\hat{O}_{\text{loc}}$ is contained within a ball of radius $r_0$,
and choose $\Lambda_{R}$ to be a larger, concentric ball of radius $r_0+R$.
This construction creates a buffer zone of linear size $R$ between the support of $\hat{O}_{\text{loc}}$ and the region where any operator $B$ can be defined.

Applying the commutator bound in Eq.~\eqref{eq:commutator_bound}, we find that the condition of the theorem is met with the parameter $\epsilon$ given by:
\begin{align}
    \epsilon(R) = C \|\hat{O}_{\text{loc}}\| e^{-\min(a, \mu/v) \cdot R}.
\end{align}
The theorem then guarantees that $\hat{L}_{\text{loc}}$ can be approximated by a strictly local operator,
$\hat{L}_{\Lambda_R}$, supported on $\Lambda_R$. The error of this approximation is bounded by $2\epsilon(R)$, leading to the final result:
\begin{align}\label{eq:err_single_term}
    \|\hat{L}_{\text{loc}} - \hat{L}_{\Lambda_R}\| \le 2 C\|\hat{O}_{\text{loc}}\| e^{-\min(a, \mu/v) \cdot R}.
\end{align}

We now consider the approximation for the full operator $\hat{L}=\sum_i \hat{L}_{i}$.
For a finite system, the cumulative error of the term-by-term approximation is bounded by a sum of exponentially decaying terms,
and is thus itself exponentially small in the buffer radius $R$.
This exponential decay easily overcomes any polynomial growth in the number of sites, ensuring the overall approximation remains efficient~\footnote{
    For an infinite lattice, $\hat{L}$ is an extensive quantity.
    For such operators, the relevant metric is not the total error but the error per site.
    Our construction, which replaces each $\hat{L}_{i}$
    with its local approximation, provides a concrete scheme that guarantees the error per site is exponentially small.
}.

In summary, by demonstrating that each term $\hat{L}_i$ admits a local approximation with an exponentially small error,
we confirm that the locality structure of $\hat{O}$ is inherited by $\hat{L}$ in the sense that each contribution admits a finite‐support approximation with exponentially small error.

\end{document}